\documentclass[%
aps,
prd,
amssymb,
amsmath,
amsfonts,
eqsecnum,
showpacs,
nofootinbib,
floatfix,
twocolumn,
twoside,
a4paper,
superscriptaddress,
longbibliography
]{revtex4-2}

\usepackage[T3,T1]{fontenc}
\usepackage{mathrsfs} % For \mathscr fonts
\usepackage{bm} % For \bm fonts
\makeatletter
\@ifclassloaded{beamer}
  { % if this is a beamer document, use sans-serif fonts
    \typeout{UsePackages: Detected beamer}
    \usepackage{tgheros}
    
  }
  { % otherwise:
    \typeout{UsePackages: Did not detect beamer}
    \ifx\asybeamer\undefined % use times for articles
    \typeout{UsePackages: Detected article}
    \usepackage[varg]{txfonts} % For math
    \usepackage{tgtermes} % For text
    \else % use beamer fonts for asy documents with beamer
    \typeout{Fonts: Detected asy for beamer}
    \usepackage{tgheros}
    
    \fi
  }
\usepackage{microtype} % For nicer alignment of text
\def\MT@register@subst@font{
  \MT@exp@one@n\MT@in@clist\font@name\MT@font@list
  \ifMT@inlist@\else\xdef\MT@font@list{\MT@font@list\font@name,}\fi}
\makeatother

\usepackage{graphicx} %
\usepackage[x11names,svgnames,rgb]{xcolor} %
\usepackage{xspace}
\usepackage{braket}
\usepackage{accents}
\usepackage{siunitx}
\usepackage{mathtools}
\DeclareSymbolFontAlphabet{\mathrm}{operators}

\definecolor{CiteColor}{rgb}{0.18039, 0.18824, 0.57255}
\definecolor{UrlColor} {rgb}{0.741, 0.173, 0.000}
\definecolor{DarkUrlColor} {rgb}{0.500, 0.110, 0.000}
\definecolor{LinkColor}{rgb}{0.25098, 0.47843, 0.04706}
\makeatletter %
\newcommand{\ShowFont}{%
  \typeout{The main font is \f@encoding \space \f@family \space %
    \f@series \space \f@shape \space at \f@size pt.}%
  \typeout{The math font sizes are \tf@size pt (main), \sf@size pt %
    (script), and \ssf@size pt (scriptscript).}%
  \typeout{The linewidth is \the\linewidth}} %
\makeatother %
\usepackage{xargs}

\usepackage{graphicx} % Include figure files
\usepackage{dcolumn} % Align table columns on decimal point
\usepackage{bm} % bold math
\usepackage{tabularx}
\usepackage{multirow}
\usepackage{capt-of}
\usepackage{color}
\usepackage{url}
\usepackage[utf8]{inputenc}
\usepackage{placeins}
\usepackage{comment}
\usepackage{enumerate} % enumerate using i)
\usepackage{cancel} % cross math

\usepackage{array}

\usepackage{natbib}
\usepackage{soul}
\usepackage{tikz-cd}

\usepackage[nolist,nohyperlinks]{acronym}
\usepackage{xspace}
\usepackage[colorlinks]{hyperref}

\usepackage[caption=false]{subfig}

\DeclareMathAlphabet{\mathbfsf}{\encodingdefault}{\sfdefault}{bx}{sl}

\newcommand{\Msun}{M_\odot}

\renewcommand{\d}{\mathrm{d}}

\newcommand{\Sone}{\bm{S_1}}
\newcommand{\Stwo}{\bm{S_2}}
\newcommand{\Lvec}{{\bm{L}}}
\newcommand{\nvec}{{\bm{n}}}
\newcommand{\Jvec}{{\bm{J}}}
\newcommand{\Lhat}{{\hat{\bm{L}}}}
\newcommand{\nhat}{{\hat{\bm{n}}}}
\newcommand{\Lrefhat}{\hat{\bm{L}}_{\mathrm{ref}}}
\newcommand{\nrefhat}{\hat{\bm{n}}_{\mathrm{ref}}}
\newcommand{\rone}{\bm{r_1}}
\newcommand{\rtwo}{\bm{r_2}}
\newcommand{\tref}{t_\mathrm{ref}}
\newcommand{\omegaref}{\mathrm{\omega_{ref}}}
\newcommand{\xhat}{{\hat{\bm{x}}}}
\newcommand{\yhat}{{\hat{\bm{y}}}}
\newcommand{\zhat}{{\hat{\bm{z}}}}
\newcommand{\chiivec}{\bm{\chi_i}}
\newcommand{\chionevec}{\bm{\chi_1}}
\newcommand{\chitwovec}{\bm{\chi_2}}
\newcommand{\omegaorb}{\omega_\mathrm{orb}}

\newcommand{\Qhat}{{\hat{\bm{Q}}}}
\newcommand{\Qvec}{{\bm{Q}}}
\newcommand{\LNhat}{\hat{\bm{L}}_{\bm{N}}}
\newcommand{\LNvec}{\bm{L}_{\bm{N}}}
\newcommand{\phiorb}{\phi_\mathrm{orb}}

\newcommand{\rvec}{\bm{r}}
\newcommand{\rhat}{{\hat{\bm{r}}}}

\newcommand{\alphaA}{\alpha_\mathrm{A}}
\newcommand{\betaA}{\beta_\mathrm{A}}
\newcommand{\gammaA}{\gamma_\mathrm{A}}
\newcommand{\alphaB}{\alpha_\mathrm{B}}
\newcommand{\betaB}{\beta_\mathrm{B}}
\newcommand{\gammaB}{\gamma_\mathrm{B}}

\newcommand{\hA}{h_\mathrm{A}}
\newcommand{\hB}{h_\mathrm{B}}

\renewcommand{\Re}[1]{\mathrm{Re}\ #1}
\renewcommand{\Im}[1]{\mathrm{Im}\ #1}

\newcommand{\aA}{\alpha_\mathrm{A}}
\newcommand{\bA}{\beta_\mathrm{A}}
\newcommand{\gA}{\gamma_\mathrm{A}}
\newcommand{\aB}{\alpha_\mathrm{B}}
\newcommand{\bB}{\beta_\mathrm{B}}
\newcommand{\gB}{\gamma_\mathrm{B}}
\newcommand{\dg}{\delta\gamma}

\newcommand{\xohat}{\hat{\bm{x}}_{0}}
\newcommand{\yohat}{\hat{\bm{y}}_{0}}
\newcommand{\xoAhat}{\hat{\bm{x}}_{0,A}}
\newcommand{\yoAhat}{\hat{\bm{y}}_{0,A}}
\newcommand{\zAhat}{\hat{\bm{z}}_{0,A}}
\newcommand{\xoBhat}{\hat{\bm{x}}_{0,B}}
\newcommand{\yoBhat}{\hat{\bm{y}}_{0,B}}
\newcommand{\zBhat}{\hat{\bm{z}}_{0,B}}

\definecolor{dodgerblue}{HTML}{1E90FF}
\definecolor{viennared}{HTML}{DA0A14}

\AtBeginDocument{%
  \hypersetup{
    citecolor=dodgerblue,
    linkcolor=dodgerblue,   
    urlcolor=dodgerblue}}

\newcommand{\UIB}{Departament de F\'isica, Universitat de les Illes Balears, IAC3 -- IEEC, Crta. Valldemossa km 7.5, E-07122 Palma, Spain}
\newcommand{\ICE}
{Institut de Ci\`encies de l'Espai (ICE, CSIC), Campus UAB, Carrer de Can Magrans s/n, 08193 Cerdanyola del Vall\`es, Spain}
\newcommand{\AEI}{Max Planck Institut für Gravitationsphysik (Albert Einstein Institut), Am M\"uhlenberg 1, Potsdam, Germany}

\allowdisplaybreaks[1]

\begin{document}

\title[Building precessing hybrids]
{Hybrid waveforms for precessing quasi-circular binary systems}

\author{Joan Llobera-Querol} \affiliation{\UIB}
\author{Sascha Husa} \affiliation{\ICE} \affiliation{\UIB}
\author{{Maria de Lluc} {Planas}} \affiliation{\AEI} \affiliation{\UIB}

\date{\today}

\begin{abstract}
The demand for long and accurate gravitational waveforms is increasing as we prepare for the next generation of detectors and seek to improve current waveform models.
However, numerical relativity waveforms, while highly accurate, are often too short for these applications due to their high computational cost.
Hybrid waveforms, which stitch together gravitational wave signals from different modeling approaches, provide a way to generate complete inspiral-merger-ringdown signals.
While hybridization is well-established for aligned-spin systems, precession introduces additional complexities due to gauge ambiguities, frame dependence, or spin dynamics.
Here we study the challenges associated with alignment of precessing waveforms and present a systematic approach for constructing hybrid waveforms of precessing quasi-circular systems.
Our approach relies on minimal assumptions about the merger waveforms and employs the quadrupole-aligned frame to mitigate mode-mixing.
Our method is designed to be robust and broadly applicable, imposing minimal constraints on the input waveforms.
This framework expands the applicability of hybridization techniques, facilitating flexible hybrid construction for parameter estimation, model calibration, and gravitational-wave data analysis.
\end{abstract}


\maketitle

\section{Introduction} \label{sec:introduction}
% OVERVIEW AND ROLE OF HYBRIDS
Accurate template waveforms are essential for gravitational wave (GW) data analysis of compact binary coalescences (CBCs). Ideally, these waveforms span the entire inspiral, merger, and ringdown (IMR) phases. 
The inspiral is well described by post-Newtonian (PN) approximations \cite{Futamase:2007zz, Blanchet:2013haa, Blanchet:2023bwj, Blanchet:2023sbv, Cunningham:2024dog, Favata:2008yd, Henry:2022ccf, Cho:2022syn, Marsat:2014xea},
or effective-one-body (EOB) resummations \cite{Buonanno:1998gg, Buonanno:2000ef, Damour:2001tu, Buonanno:2002fy, Pompili:2023tna, Ramos-Buades:2023ehm, vandeMeent:2023ols, Khalil:2023kep, Nagar:2020pcj, Riemenschneider:2021ppj, Gamba:2021ydi},
but PN methods break down in the late inspiral as orbital velocities increase \cite{Blanchet:2013haa, Buonanno:1998gg}, requiring numerical solutions of the full field equations.
For large mass ratios, self-force techniques have recently achieved high-accuracy inspirals even for comparable-mass binaries \cite{LeTiec:2013uey, Albertini:2022rfe, Wardell:2021fyy}.
Hybrid waveforms, which synthesize solutions from different approximation methods, have proven fruitful, particularly since the 2005 breakthrough in numerical relativity (NR) \cite{Pretorius:2005gq}.
They are widely used to inform and validate waveform models \cite{Husa:2015iqa, Varma:2018mmi} and will become even more useful to describe the very long waveforms that will be observed with third-generation and space-based detectors \cite{LISAConsortiumWaveformWorkingGroup:2023arg}.

% HYBRIDIZATION IN TWO SENTENCES
Hybridization involves selecting two physically equivalent waveforms generated by different methods, aligning them in time and orientation, and smoothly combining them.
Beyond producing a hybrid, this waveform comparison is valuable in itself, as it quantifies discrepancies between the waveforms in an overlapping region.

% HYBRIDS IN QUASI-CIRCULAR AND PRECESSION ISSUES
For quasi-circular (QC) binaries with aligned spins, hybrid waveform construction has become relatively routine \cite{Santamaria:2010yb, Hannam:2010ky, Ajith:2007qp, Ajith:2007kx, Ajith:2007xh, Ajith:2012az, Boyle:2009dg, MacDonald:2011ne, Boyle:2011dy, Damour:2010zb, CalderonBustillo:2015lrg}.
Here, we extend these methods to precessing binaries, where spin directions evolve over time, causing the orbital plane to precess \cite{Apostolatos:1994mx,Kidder:1995zr}.
This modulation complicates waveform modeling \cite{Schmidt:2010it}, introducing ambiguities in spin parameterization and frame selection.
In particular we face the challenge that when the waveform parameters are time dependent, and defined in terms of strong-field properties (spins associated with the black hole (BH) horizons), it becomes non-trivial to relate the spin evolution to the evolving waveform, which is defined far away from the BHs, and indeed ideally at null infinity. 
Parameterizing the waveform directly in terms of properties of the waveform would be desirable, but it is not clear to which degree this is feasible in practice.
Additionally, spin directions are gauge-dependent in the strong field region, and different modeling approaches, such as PN and NR, often use distinct coordinate gauges \cite{Mitman:2021xkq, Boyle:2019kee}.

% REFERENCE FRAME: PRECESSING CHALLENGES
Choosing an appropriate reference frame simplifies the description of the orbital geometry, spin vectors, and GW signal. This frame can be either inertial or time-dependent.
In the presence of orthogonal spin components, orbital precession prevents the definition of a natural inertial frame.
This motion modulates waveform modes in any inertial frame, coupling modes with the same $\ell$ but different $m$-- we refer to this phenomenon as \emph{mode-mixing}.
As a result, mode representations vary between frames, much like spin vector descriptions.
This inherent complexity makes constructing waveform models for precessing systems particularly challenging.
However, this challenge can be addressed by employing non-inertial frames that co-rotate with the orbital plane.

% IN THIS WORK...
In this work, we present and compare two existing methods for hybridizing precessing waveforms, addressing their alignment and comparison in both an inertial frame and a time-dependent coprecessing frame. Describing waveforms in the latter simplifies the task, since then waveforms resemble those of non-precessing systems.
We assess each method's strengths and limitations, and justify our preference for the coprecessing frame approach.
Additionally, we account for the constraints of current waveform models and NR simulations, outlining how we navigate the ambiguities and tradeoffs inherent to hybridization.

% SUMMARY
In this paper we do not seek to generate optimized hybrid waveforms for specific systems, but rather focus on discussing the challenges of aligning precessing waveforms and of developing a robust and generalizable methodology for hybrid construction, assuming minimal information about the details of how the ``inspiral'' and ``merger'' waveforms are constructed. In order to simplify our hybrid construction we neglect several issues that would further extend the scope of this paper and which we leave for future work.
% BBH SYSTEMS NOTATION
First,  we restrict ourselves to hybridizing QC binary black hole (BBH) waveforms.
These systems are characterized by an eight-dimensional parameter space, defined by two component masses $m_i$ and two dimensionless spin vectors $\chiivec$ with magnitudes $\chi_i$.
We neglect the small changes of the BH masses and spins during the inspiral, and describe them using the mass ratio $q = m_1/m_2 \geq 1$, while the total mass $M = m_1 + m_2$ serves as a scale parameter. We expect our methods to also broadly apply to binary systems with neutron stars, where however one needs to account for the expanded parameter space with tidal deformabilities $\Lambda_i$ and the object's individual masses.

% CHALLENGES OF ALIGNMENT
In addition to general modeling inaccuracies, waveforms will also typically differ in parameterization, in particular due to different gauges, or e.g.~initial transients in NR simulations. A particular problem for precessing systems is that the spin directions will be gauge dependent, see e.g. \cite{Campanelli:2006fy,Szabados:2009eka}~or~\cite{Boyle:2019kee} for definitions in the NR context. Such ambiguities will in general become more pronounced as the binary shrinks toward the merger. A previous study has found spin angle differences between a PN description and NR simulations performed in a generalized harmonic gauge not to be larger than a few degrees \cite{Ossokine:2015vda},
however a more thorough understanding covering further gauges and a larger parameter space will be required in the future. 
For a discussion of choosing the consistent Bondi-Metzner-Sachs (BMS) frames for the waveforms to be compared or hybridized see \cite{Sun:2024kmv},
where waveforms are computed at null infinity with Cauchy-Characteristic extraction \cite{Winicour:2008vpn,Moxon:2021gbv}.

% BMS FRAME
One option to address parameterization ambiguities is to 
optimize the choice of inspiral (specifically PN) waveform that is glued to the merger description over the PN waveforms' intrinsic parameters, as well as the extrinsic degrees of freedom corresponding to time shift and rotations, which we use here. 
While the work \cite{Sun:2024kmv} optimizes over all intrinsic parameters in the same way, we choose not to perform such an optimization here, and leave a study of which parameters should be optimized over, and which ones to keep fixed, e.g. to avoid overfitting to degeneracies, to future work.

Our proposed method addresses the challenges discussed above while remaining broadly applicable. Consequently, we do not prescribe specific criteria, such as the required merger waveform length for successful hybridization.

% PAPER STRUCTURE
Section~\ref{sec:conventions} introduces the conventions and reference frames relevant to precessing binaries, while in~Sec.~\ref{sec:nonprecessing_hybrids} we briefly sketch the established methods for constructing multi-modal non-precessing hybrid waveforms, providing context for the precessing case.
In~Sec.~\ref{sec:building_hybrids} we detail our methods and the key challenges of precessing hybridization.
Finally, we discuss our results in Sec.~\ref{sec:discussion}.

Throughout this paper we use geometric units with ${G=c=1}$.

\section{The phenomenology of precessing waveforms} \label{sec:conventions}
% 2A. CBC WAVEFORMS
\subsection{Spherical harmonic decomposition of the waveform} \label{ssec:waveform_modes}

% SPHERICAL HARMONICS
The GW signal from a compact binary system depends on its intrinsic parameters $\bm{\lambda}$, and the emission direction $\left(d_L,\theta,\varphi\right)$.
It is typically decomposed into spherical harmonics of spin weight -2,
\begin{equation} \label{eq:modes}
    h\left(t, \bm{\lambda}, \bm{d_L}\right) = \dfrac{1}{d_L}\ \sum_{\ell\geq 2} \sum_{m=-\ell}^{\ell} h_{\ell m} \left(t, \bm{\lambda}\right)\cdot {}^{-2}\mathcal{Y}_{\ell m}\left(\theta,\varphi\right)
\end{equation}
where $d_L$ is the luminosity distance to the source of the perturbation and ${}^{-2}\mathcal{Y}_{\ell m}\left(\theta,\varphi\right)$ is the basis of spherical harmonics of spin weight $-2$.

% GRAVITATIONAL WAVE MODES
Each mode $h_{\ell m}(t)$ is a complex-valued function that can be further broken down to its amplitude $A_{\ell m}(t)$ and phase $\varphi_{\ell m}(t)$ components, as described by
\begin{equation}
    h_{\ell m}(t) = A_{\ell m}(t)\ \mathrm{e}^{i \varphi_{\ell m}(t)}.
\end{equation}

For non-precessing systems these modes simplify due to the symmetry about the orbital angular momentum, which relates the positive and negative $m$ modes as
\begin{equation}
    h_{\ell\, -m} = (-1)^m\ h_{\ell m}^\ast
\end{equation}
with the dominant contribution coming from the $(\ell,|m|) = (2,2)$ modes.
The frequency associated with each mode is given by $f_{\ell m} = \dot{\varphi}_{\ell m}$, and for non-precessing systems during the inspiral, these frequencies approximately satisfy
\begin{equation} \label{eq:modes_frequency}
    f_{\ell m} \approx \dfrac{m}{2} f_{22}.
\end{equation}

% DIFFERENT TIME PARAMETERIZATIONS
Establishing a precise correspondence between the dynamical evolution of the binary (i.e. the trajectory of masses and spins evolution) and the emitted GWs requires careful consideration of different time parameterizations. Due to the travel time of gravitational radiation, a time lag arises between the spin evolution and the observed waveform, further complicated by gauge effects that make an exact mapping between these time coordinates nontrivial \cite{Hamilton:2018fxk}.
%Different waveform models and NR catalogs address this challenge in distinct ways.

% ORBITAL AND WAVEFORM FREQUENCIES
Despite these complications, for non-precessing systems an approximate relation holds between the frequency of the dominant modes and the orbital frequency,
\begin{equation} \label{eq:f22-forb}
    f_{22} \simeq 2 f_{orb},
\end{equation} with high accuracy (see e.g.~\cite{Estelles:2020twz}), that also applies to orbit-averaged frequencies for eccentric systems. This provides a practical way to link the system's dynamics with the GW evolution, while circumventing the difficulties posed by non-matching time parameterizations.

% ECCENTRICITY
However, NR simulations are only approximately QC and often retain small residual eccentricity, which can introduce minor deviations from the QC behavior.

% 2B. DESCRIBING CBC SYSTEMS
\subsection{Inertial reference frames} \label{ssec:reference_frames}

% Lref FRAME
A widely used convention in gravitational waveform modeling is the \emph{LAL source-frame}, following the conventions established by the LALSuite software library \cite{Schmidt:2017btt,lalsuite}.
In this framework, the $\zhat$-axis aligns with $\Lrefhat$, the $\xhat$-axis with $\nrefhat$, and $\yhat$ completes the triad as $\zhat \times \xhat$. 
$\Lvec$ is the orbital angular momentum of the binary, and $\nvec=\rone-\rtwo$ points from the lighter to the heavier object.
The ``ref'' subscript indicates that $\Lhat,\,\nhat$ are defined at a reference point of the binary's evolution, specified by a time $t=\tref$ or parameters like orbital frequency $\omega=\omegaref$ or separation $r\equiv|\nvec|=r_\mathrm{ref}$.
For practical purposes, this frame is sometimes defined with the Newtonian angular momentum $\LNvec$ instead of $\Lvec$ defining the $\zhat$-axis.

% L0 and J FRAMES
The \emph{LAL source-frame}, also called the $L_\mathrm{ref}$-frame for the vector that defines its $\zhat$-axis, coexists with other commonly used inertial frames, such as the $L_0$-frame and the $J$-frame. The subscript ``0'' refers to the initial point of evolution, while $\Jvec$ represents the total angular momentum $\Jvec = \Lvec + \Sone + \Stwo$. 
Due to the (approximate) conservation of $\Jvec$, all the $J_\mathrm{ref}$-frames are (approximately) equivalent for any choice of reference point, except near the merger, which is why the ``ref'' subscript is typically omitted when discussing the J-frame.

\begin{figure}   
    \includegraphics[width=\columnwidth]{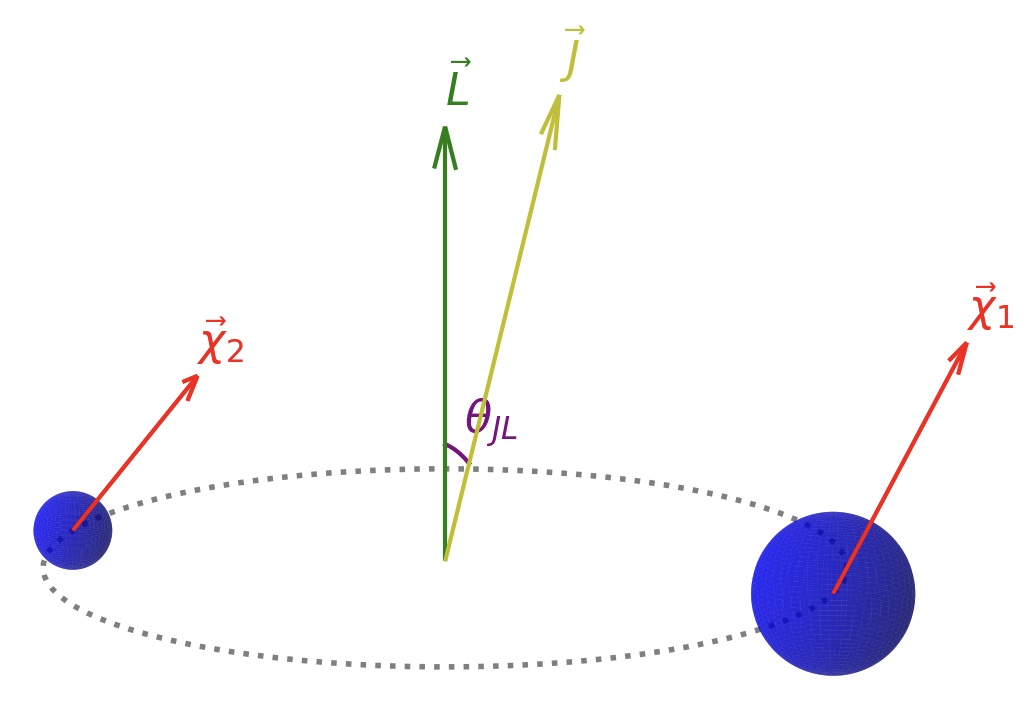}
    \caption{Illustration of a binary system with its key vector quantities: $\chionevec$, $\chitwovec$, $\Lvec$, and $\Jvec$. Different frame choices offer advantages in various contexts. In frames where $\zhat\ \parallel \Lvec$, the x-y plane moves to match the orbital plane.}
    \label{fig:binary_vectors}
\end{figure}

These frames offer alternative descriptions of the spatial orientation and spin components of the binary system, with distinct advantages in different contexts. Figure~\ref{fig:binary_vectors} shows a representation of a binary system with the spins and angular momentum that illustrates the discussion.

% RELATION BETWEEN FRAMES
To relate two frames, define the matrix $\bm{R}_\mathrm{AB}$ 
where its columns are the basis vectors of frame B expressed in the frame A.
Then, vectors like spins transform as $\bm{v}_\mathrm{A} = \bm{R}_\mathrm{AB}\ \bm{v}_\mathrm{B}$, where the subscript indicates the frame in which it is expressed.
Any of such frame transformations can be decomposed into three intrinsic axis rotations, for example, in the $z-y-z$ convention.
In this case, $\bm{R}_\mathrm{AB}$ consists of a rotation by $\alpha$ around the z-axis, followed by $\beta$ around the new y-axis, and finally $\gamma$ around the new $z$-axis:
\begin{equation} \label{eq:rotation_decomposition}
    \bm{R}\left(\alpha,\,\beta,\,\gamma\right) = \mathcal{R}_z(\alpha)\ \mathcal{R}_y(\beta)\ \mathcal{R}_z(\gamma)
\end{equation} where $\mathcal{R}_i(\theta)$ is the active 3D rotation matrix by an angle $\theta$ around axis $i$.
Hence, the Euler angles $\alpha,\,\beta,\,\gamma$ fully describe~$\bm{R}_\mathrm{AB}$.

% FIGURE: BINARY WITH VECTORS

% RELATION BETWEEN WAVEFORMS
Waveforms from different frames are related by:
\begin{equation} \label{eq:waveform_frames}
    h^\mathrm{B}_{\ell m} \left( t, \bm{\lambda} \right) = \sum_{m'=-\ell}^\ell h^\mathrm{A}_{\ell m'} \left( t, \bm{\lambda} \right)\ \mathcal{D}^{\ell}_{m\, m'} \left( \alpha,\,\beta,\,\gamma \right),
\end{equation}
where $\mathcal{D}^\ell_{m\, m'} \left( \alpha,\,\beta,\,\gamma \right)$ are the Wigner D-matrices (see \cite{Pratten:2020ceb} for conventions).

% 2C. PHENOMENOLOGY OF PRECESSING SYSTEMS
\subsection{Precessing systems} \label{ssec:precession}

In aligned spin systems the component spins are parallel or anti-parallel to the orbital angular momentum $\Lvec$. In this case, directions of the component spins and of the orbital plane (and thus of $\Lvec$) are preserved, and the spacetime exhibits equatorial symmetry. Such systems can then be described by three dimensionless intrinsic parameters, the mass ratio and dimensionless spin projections along $\Lvec$.
When the spins are misaligned with $\Lvec$, the spin directions, and in general also the orbital plane and $\Lvec$ become time-dependent, and the equatorial symmetry is lost. For misaligned binaries one then needs seven instead of three intrinsic parameters: the mass ratio and two spin vectors.
The tightening of the binary is primarily driven by the ``aligned'' spin components, whereas the precession of the spins and the orbital plane is primarily driven by the spin components orthogonal to $\Lvec$. This phenomenology is simplified by the fact that the magnitude of the spin components parallel and orthogonal to $\Lvec$ remain approximately constant during the inspiral \cite{Kidder:1995zr,Schmidt:2010it,Schmidt:2014iyl}.
Differences in spin or orbital angular momentum between waveform models are typically limited to a few degrees \cite{Ossokine:2015vda}.

% PRECESSING SYSTEMS
In the aligned spin case, the conservation of the orbital plane defines a natural inertial reference frame, simplifying waveform modeling. In contrast, precessing systems lack this simplicity, and a careful choice of reference frame is required.
% CONSEQUENCES
In an inertial frame, waveform modes are a combination of the modes in the coprecessing frame, exhibiting \emph{mode-mixing}
according to Eq.~\eqref{eq:waveform_frames}.
As a result, several key properties are altered, such as the dominance of the (2,|2|) modes and the relationships between mode frequencies in Eq.~\eqref{eq:modes_frequency}.

% 2D. THE COPRECESSING FRAME
\subsection{Coprecessing frame} \label{ssec:coprecessing_frame}

% NON-INERTIAL FRAMES
Waveform modeling of precessing systems can be simplified by using non-inertial frames that co-rotate with the orbital plane.
A \emph{coprecessing frame} has its $\zhat$-axis aligned with the orbital angular momentum $\Lhat$ at all times \cite{Schmidt:2010it}.
If the frame also aligns the $\xhat$-axis with $\nhat$, it becomes a \emph{co-orbital frame}, which matches the \emph{LAL source-frame} at reference time. In this frame, the waveform's oscillatory nature is absent, further simplifying the modeling.

% QUADRUPOLE-ALIGNED FRAME
Similarly, the \emph{quadrupole-aligned (QA) frame} is a non-inertial frame where the $\zhat$-axis aligns with the direction that maximizes the amplitude of the quadrupole emission, denoted as $\Qhat$. Since this direction roughly coincides with the orbital angular momentum ($\Qhat \simeq \Lhat$) \cite{Schmidt:2010it}, the QA frame behaves similarly to a coprecessing frame. The orbital plane rotation is fixed using the minimal rotation condition \cite{Boyle:2011gg}.
In addition to $\Lvec$ and $\Qvec$, the Newtonian orbital angular momentum $\LNvec \equiv \mu\,\rvec\times\dot{\rvec}$, can also define the $\zhat$-axis. However, while $\LNhat$ exhibits nutating behavior, $\Lhat$ and $\Qhat$ evolve more smoothly \cite{Schmidt:2010it}.

% TWISTING-UP
The orbital timescale is much shorter than the precession timescale, allowing us to neglect the power radiated due to precession in the inspiral phase.
Consequently, the waveform in the QA frame closely matches that of a non-precessing system with the same aligned spin components ($h^{QA}\left(t,\bm{\lambda}\right) \approx h^{AS}\left(t,\bm{\lambda}\right)$) \cite{Schmidt:2010it}, in what is known as the \emph{twisting-up approximation}, crucial for modeling precessing systems (see \cite{Estelles:2021gvs, Ramos-Buades:2023ehm} among others).

% MODE MIXING
To highlight the absence of mode-mixing in the QA frame, we note that subdominant modes have amplitudes orders of magnitude lower than the $(2,|2|)$ modes, and their frequencies satisfy Eq.~\eqref{eq:modes_frequency}. In contrast, in an inertial frame, subdominant modes evolve irregularly, while in the QA frame, their evolution is monotonic, ensuring that all physical information is captured by the QA frame modes.

However, the observed signal in an inertial frame is still influenced by time-dependent rotation effects and reconstructing the full waveform requires an accurate description of both the non-precessing modes and the precession-induced modulations.

% TWISTING-UP AND EULER ANGLES
The rotation $\bm{R}$ from the QA frame to an inertial frame is time-dependent, as are the angles $\alpha(t)$, $\beta(t)$, and $\gamma(t)$ that define it.
Since the $\zhat$-axis in the QA frame aligns with $\Qhat$, $\alpha$ and $\beta$ are given by the components of $\Qhat$:
\begin{align} \label{eq:alpha}
    \alpha &= \arctan{\left( \dfrac{\hat{\mathrm{Q}}_y}{\hat{\mathrm{Q}}_x}\right)} \approx \arctan{\left( \dfrac{\mathrm{L}_y}{\mathrm{L}_x}\right)}, \\  \label{eq:beta}
    \beta &= \arccos{\left( \hat{\mathrm{Q}}_z \right)} \approx \arccos{\left( \dfrac{\mathrm{L}_z}{\mathrm{L}} \right)}.
\end{align}
The third angle, $\gamma$, satisfies the minimal rotation condition \cite{Buonanno:2002fy,Boyle:2011gg}
\begin{equation} \label{eq:MRC}
    \dot{\gamma} = -\dot{\alpha}\, \cos{\beta},
\end{equation}
parameterizing the choice of $\xhat$ and $\yhat$ within the orbital plane.

This condition ensures that the orbital plane evolves naturally with the binary's rotation, i.e. the definition of $\xhat,\,\yhat$ does not introduce an artificial phase. When the condition is met, $\omegaorb$ accurately captures the binary's motion on the (precessing) orbital plane.

The coprecessing frame is then defined by
\begin{align}
    \xhat &= \begin{pmatrix}
        \cos{\alpha}\cos{\beta}\cos{\gamma} - \sin{\alpha}\sin{\gamma} \\
        \sin{\alpha}\cos{\beta}\cos{\gamma} + \cos{\alpha}\sin{\gamma} \\
        -\sin{\beta}\cos{\gamma}
    \end{pmatrix} \label{eq:xaxis} \\
    \yhat &= \begin{pmatrix}
        -\cos{\alpha}\cos{\beta}\sin{\gamma} - \sin{\alpha}\cos{\gamma} \\
        -\sin{\alpha}\cos{\beta}\sin{\gamma} + \cos{\alpha}\cos{\gamma} \\
        \sin{\beta}\sin{\gamma}
    \end{pmatrix} \label{eq:yaxis} \\
    \zhat &= \begin{pmatrix}
        \cos{\alpha}\sin{\beta} \\
        \sin{\alpha}\sin{\beta} \\
        \cos{\beta}
    \end{pmatrix} \label{eq:zaxis}.
\end{align}
Here, the $z$-axis matches $\Qhat$, and $\gamma$ parameterizes the choice of $\hat{x}$ and $\hat{y}$ within the orbital plane.

The rotation matrix
\begin{equation} \label{eq:rotationmatrix}
    \bm{R} = \begin{pmatrix} \xhat\ \ \yhat\ \ \zhat \end{pmatrix}
\end{equation} is coherent with Eq.~\eqref{eq:rotation_decomposition} and transforms vectors from the QA frame to an inertial frame:
\begin{equation} \label{eq:R_in_vectors}
    \bm{v}_\mathrm{inertial} = \bm{R} \cdot \bm{v}_\mathrm{QA}, \qquad \bm{v}_\mathrm{QA} = \bm{R}^\mathrm{T} \cdot \bm{v}_\mathrm{inertial}
\end{equation}
where the subscripts indicate coordinates in the inertial and QA frames.

The parameterization of the coprecessing frame in terms of Euler angles is intuitive: $\beta$ specifies the inclination of $\Lhat$ with respect to the $\zhat$ axis, while $\alpha$ expresses its rate of precession. While alternative representations of the rotation group $\mathbb{SO}$(3), such as quaternions, eliminate singularities near $\beta \approx 0$ (i.e. when the rotation is close to the identity) without affecting the results, Euler angles are retained for their intuitive interpretation, and the associated singularities can be handled without difficulty (see Sec.~\ref{ssec:coprecessing_frame}).

% IMPLEMENT THE QA FRAME
Two methods for implementing the QA frame have been proposed: one maximizes the magnitude of $(2,|2|)$ modes through rotations \cite{Schmidt:2010it, Schmidt:2012rh}, and the other aligns with the principal axis of the $\langle \mathcal{L}_{(ab)} \rangle$ tensor \cite{OShaughnessy:2011pmr}.
Both approaches are shown to be equivalent \cite{Boyle:2011gg}, with a succinct review in App.~\ref{app:ObtainingQ}.
In both cases, the QA frame is defined solely using the waveform, which is crucial for our method when searching for gauge-independent quantities.

\section{Non-precessing hybrids} \label{sec:nonprecessing_hybrids}
This section briefly reviews the construction of non-precessing hybrids with two goals: establishing the notation used throughout and laying the groundwork for the methodology we will later extend to precessing systems.

% ALIGNMENT IS ROUTINE
As discussed in Sec.~\ref{sec:conventions}, non-precessing systems have a natural preferred frame in which the waveform is described, eliminating the need for frame alignment, and spin directions remain manifestly constant during the evolution by symmetry. 
As in this work we are only concerned about QC systems, we can also neglect the evolution of orbital eccentricity. For simplicity here we neglect the small changes in the masses and spin magnitudes due to tidal heating and tidal torquing (see however
\cite{Mukherjee:2023pge} for a recent study that does take into account tidal heating of BHs). We thus consider all the intrinsic parameters that describe the system as constant during the inspiral, which greatly simplifies the hybrid construction. Our discussion of hybridizing the waveform will focus on the GW strain, although other choices are possible, see e.g.~\cite{Hannam:2010ky} for using $\Psi_4=\ddot{h}$. Also, it could be of interest to apply the hybridization procedure to other quantities, such as the energy or angular momentum flux or the orbital motion, which we leave for future work.

% NOTATION
We begin the hybridization with a waveform describing the merger and ringdown, denoted as ``A'' with strain~$\hA$, referred to as the \emph{merger description}.
To extend it into the inspiral regime, we use a second waveform ``B'', with strain~$\hB$, called the \emph{inspiral description}.

% t_0, phi_0 and psi_0
Alignment thus reduces to determining three shift parameters, corresponding to the degrees of freedom available for adjusting the relative positioning of the waveforms: a time shift~$t_0$, a phase shift~$\varphi_0$ modifying each $(\ell, m)$ mode by $m\,\varphi_0$ at leading order \cite{Arun:2008kb},
and a polarization phase shift~$\psi_0$, accounting for differences in conventions, which affects all modes equally and is constrained to two standard values $\psi_0\in{0,\pi}$, corresponding to the two polarizations that are consistent with the equatorial symmetry of non-precessing binaries (for details see the discussion in \cite{CalderonBustillo:2015lrg}).

% ALIGNMENT IS MINIMZATION OF DISCREPANCY
Alignment consists of choosing $t_0,\,\varphi_0,\,\psi_0$ to minimize a discrepancy measure $\delta$ quantifying the difference between $\hA$ and $\hB$.
Various choices for $\delta$ exist, reflecting different alignment strategies.
A common case is that the waveform $\hA$ is provided by a NR simulation. In this case (but possibly also when the waveform is constructed in other ways), it is typical that the waveform exhibits oscillations due to residual eccentricity. It is then advisable to measure the discrepancy over a time interval -typically using integrals- rather than a single reference point.
Early approaches, such as those in \cite{Ajith:2007qp,Ajith:2007kx,Ajith:2007xh,Ajith:2012az}, minimized the squared difference between the time-domain strains, summed over $+$~and~$\times$ polarizations, and often introducing an amplitude scaling factor.
Meanwhile, other works like \cite{Boyle:2008ge,Boyle:2009dg,MacDonald:2011ne,Boyle:2011dy} minimize the phase difference between waveforms.
Posterior works \cite{CalderonBustillo:2015lrg} define a time shift after matching the wave frequencies and a phase shift by matching the phase at one point.
Frequency-domain techniques have also been developed, with blending functions defined in terms of $f$ \cite{Santamaria:2010yb, Damour:2010zb}.

% MULTI-MODE HYBRIDS
An additional challenge is the coherent alignment of multiple waveform modes.
Higher-order modes become increasingly prominent in unequal-mass spinning systems, particularly in the presence of precession.
A common strategy for aligning multiple modes at once is to consider and match smoother quantities, such as the amplitude and frequency of each mode, rather than directly trying to align oscillatory complex modes.
This approach enhances the robustness of the alignment procedure and is widely adopted in waveform modeling.

% HYBRIDIZATION
Once the alignment is established, the hybridized waveform is typically constructed as
\begin{equation} \label{eq:blend}
    h_\textrm{hyb} = \begin{cases}
        \hB & t\leq t_1 \\
        \left[1-w(t)\right] \hB + w(t)\ \hA & t_1 < t \leq t_2 \\
        \hA & t_2 < t
    \end{cases}
\end{equation}
where $(t_1,t_2)$ is the hybridization interval and $w(t):[t_1,t_2]\to[0,1]$ is a smooth blending function satisfying $w(t_1)=0,\, w(t_2)=1$.
Choices for $w(t)$ include linear, cosine, hyperbolic tangent, and exponential functions, each with different smoothing properties.
When $t_1=t_2$, the transition occurs at a single matching point.

% CONSEQUENCES
By incorporating these alignment and hybridization strategies, we build upon past work to establish a framework for waveform hybridization in the precessing case.
The next step is to extend this methodology to precessing systems, where additional complexities arise due to the intricate orbital and spin dynamics.

\section{Building precessing hybrids} \label{sec:building_hybrids}

Hybridizing waveforms for precessing systems involves two key steps. First, ensuring that the inspiral and merger descriptions represent the same physical system, which in turn requires careful handling of frame rotations and spin dynamics. Second, achieving precise alignment between the waveforms to construct a coherent hybrid, requiring an appropriate choice of alignment frame and resolution of the challenges outlined above.

As mentioned in the introduction, ideally we would like to parameterize the waveform directly in terms of its intrinsic properties in a gauge invariant way. In absence of knowing such a parameterization we instead choose the standard approach of parameterizing the waveforms in terms of the BH masses and spins, and we work with a coprecessing frame (the \emph{LAL source frame}) defined in terms of orbital quantities and BH spins. We can then relate this frame to the QA frame that is defined in terms of the waveform as discussed in Sec.~\ref{ssec:QL}.

% MINIMUM INFORMATION FOR OUR METHOD
To develop a hybridization method that is as general and widely applicable as possible, we only require minimal information from the merger waveform we start with.
To align CBC descriptions, each description must provide both the waveform modes, $h_{\ell m}(t)$, and the spin evolution, $\chi_i(\omegaorb)$, in a given inertial frame.
If this frame corresponds to the $L_\mathrm{ref}$-frame, it can be identified by a reference frequency $\omegaref$.
Our method is independent of gauge-sensitive quantities, such as the coordinate positions of the two objects, and does not rely on additional parameters like the precession of the orbital plane $\hat{L}(\omegaorb)$.

In the alignment procedure, crucial for blending and hybridization, we choose to keep the merger waveform $\hA(t)$ unaltered, such that it is recovered in the hybrid waveform after the hybridization window.
Next, we provide an overview of the algorithm to build a hybrid starting with a merger waveform $\hA(t)$.
\begin{enumerate}
    \item Select a reference frequency $f_\mathrm{ref}$ to generate the inspiral waveform.
    \item Measure the spins at the appropriate frame to generate the inspiral waveform. Specifically, \begin{enumerate}
        \item Identify the spin values ${(\chi_i)}_A$ where the orbital frequency satisfies $\omegaorb=\pi f_\mathrm{ref}$.
        \item Compute the precession angles $\alphaA,\,\betaA,\,\gammaA$ at the time when $\langle f_{22}^\mathrm{QA} \rangle = f_\mathrm{ref}$.
        \item Transform the spins ${(\chi_i)}_A$ to the QA frame at the reference frequency.
    \end{enumerate}
    \item Use the transformed spins and reference frequency to generate the inspiral waveform $\hB(t)$ with the approximant of choice.
    \item Align the waveforms by decomposing them into the coprecessing-frame waveforms and the precession angles. Specifically, \begin{enumerate}
        \item Compute the coprecessing-frame waveform $h^\mathrm{QA}(t)$ and precession angles $\{\alpha,\beta,\gamma\}(t)$ for both $\hA$ and $\hB$.
        \item Align the coprecessing-frame waveforms by determining the optimal shifts $t_0,\,\varphi_0,\psi_0$ that minimize a chosen discrepancy measure $\delta$ over a selected time interval.
        \item Align the precession angles by finding the optimal inertial rotation $\mathbf{R}_0$ that minimizes a chosen discrepancy measure $\delta$ across the same time interval.
    \end{enumerate}
    \item Construct the hybrid waveform in the coprecessing frame by smoothly blending the inspiral and merger segments using a transition window defined by a function $w(t)$, and apply the same blending to the precession angles.
    \item Reconstruct the full inertial-frame hybrid waveform by applying the hybrid precession angles to the hybrid coprecessing-frame waveform.
\end{enumerate}

In Sec.~\ref{ssec:dof}, we provide an overview of the degrees of freedom of these systems and the need of aligning.
In Sec.~\ref{ssec:reference_point} we discuss how to determine an adequate reference point [step 1].
In Sec.~\ref{ssec:QL} we discuss the construction of the QA frame in order to generate the inspiral waveform while in Sec.~\ref{ssec:orbital_plane}, we discuss the fixing of the orbital plane rotation [step 2.c].
In Sec.~\ref{ssec:hybrids_inertial} we explore the option of aligning in the inertial frame, which we discard.
On the contrary, in Sec.~\ref{ssec:hybrids_coprecessing} we explore the alignment of the precession angles [step 4.c].
In Sec.~\ref{ssec:blending} we review the blending function [step 5].

\subsection{Degrees of freedom of aligning waveforms} \label{ssec:dof}

% DEGREES OF FREEDOM
Alignment of precessing waveforms involves dealing with the degrees of freedom that, while not physically significant, influence the waveform description:
\begin{enumerate}
    \item Three degrees of freedom --$t_0,\,\varphi_0,\,\psi_0$-- correspond to shifts in time, phase, and polarization, as explained in Sec.~\ref{sec:nonprecessing_hybrids} for non-precessing systems. These affect the waveform modes in any frame in a simple manner
    \begin{equation} \label{eq:shifts}
        h_{\ell m}(t) \overset{t_0,\,\varphi_0,\,\psi_0}{\longrightarrow} h_{\ell m}(t-t_0) \cdot \mathrm{e}^{i(m\varphi_0 + 2 \psi_0)}
    \end{equation}
    Note that while time shifts commute with frame changes (see Eq.~\eqref{eq:waveform_frames}), phase shifts do not.
    \item Three additional degrees of freedom arise from the choice of inertial frame in which the system is described. These affect both the dynamics evolution and the waveform modes in an inertial frame as described in Sec.~\ref{ssec:precession}, but not the waveform modes in the QA frame.
    Regardless of the chosen alignment frame, hybridization ultimately requires aligning the inertial frames of both waveforms. This alignment involves a 3D rotation $\bm{R}_0$, which accounts for the three degrees of freedom.
    We parameterize $\bm{R}_0$ with the Euler angles $\{\alpha_0,\,\beta_0,\,\gamma_0\}$ in the z-y-z convention. These angles should not be confused with the time-dependent Euler angles $\{ \alpha(t), \beta(t), \gamma(t) \}$ which describe the inertial frame in which the GW signal is expressed.
    Then, analogous to Eq.~\eqref{eq:shifts}, we write:
    \begin{equation}
        \{ \alpha(t), \beta(t), \gamma(t) \} \overset{\alpha_0,\,\beta_0,\,\gamma_0}{\longrightarrow} \bm{R}_0 \circ \{ \alpha(t), \beta(t), \gamma(t) \}
    \end{equation}
    where $\circ$ denotes the action of $\bm{R}_0$ on the Euler angles following standard linear algebra conventions.
\end{enumerate}

% EQUIVALENCE CLASS
Waveforms that differ only by the aforementioned six degrees of freedom --time and phase shifts, and spatial rotations-- are considered physically equivalent and belong to the same equivalence class.
In such cases, an optimal choice of these parameters can align the two waveforms exactly.
However, a waveform also depends on the intrinsic parameters of the binary --the mass ratio and spin vectors in the QC case-- as well as on the method used to generate it (e.g. NR, PN or approximants).
For precessing systems, two equivalent waveforms might have different intrinsic parameters if the spin vectors are specified at different reference times.

% REAL APPLICATION
But this is complicated even further when two waveforms are produced by different methods, since even with the same intrinsic parameters waveforms will differ due to inherent gauge choices and numerical precision limitations.
Therefore, in all practical applications, hybridization of precessing waveforms is not trivial, and the hybridization process must optimize these degrees of freedom to minimize discrepancies between descriptions as much as possible.

% NON-PRECESSING vs. PRECESSING DISCUSSION
In the non-precessing case the natural choice of axis is to set $\Lhat=\zhat$, so the only degree of freedom for choosing a specific frame corresponds to rotating  $\hat{x}$ and $\hat{y}$ within the orbital plane. For aligned-spin binaries this degree of freedom is degenerate with the binary phase shift $\varphi_0$ since a rotation around the orbital momentum axis will only impact the waveform with a mode-coherent phase shift, that can be reverted with the choice of $\varphi_0$. 
For this reason, hybridization of non-precessing waveforms does not need to consider any of the frame alignment degrees of freedom.
For precessing binaries these simplifications do not arise, in particular the phase shift $\varphi_0$ and rotations of the frame around the $\zhat$-axis are not degenerate.

\subsection{Determining a reference point} \label{ssec:reference_point}

% The hybridization point is a metaparameter.
In order to generate the inspiral waveform with the approximant of choice, spin vectors must be expressed in the appropriate frame (typically \emph{LAL source-frame}) at a specific reference point.

\begin{figure}   
    \includegraphics[width=\columnwidth]{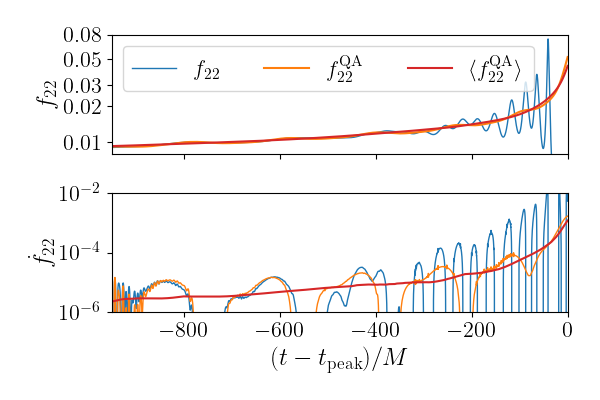}
    \caption{Comparison of the (2,2)-mode frequency in an inertial frame (blue), in the coprecessing frame (orange), and the orbit-averaged value in a coprecessing frame (red) for the NR waveform \texttt{SXS:BBH:0165}. Top panel shows frequency and bottom panel shows its first derivative, where the non-monotonicity of $f_{22}$ and $f^\mathrm{QA}_{22}$ appears evident.}
    \label{fig:fdot}
\end{figure}

The reference point in a binary's evolution should be selected unambiguously and with a gauge-independent criterion. We face the problem that we want to choose a consistent reference point for both the wave signal and the orbital and spin dynamics, which parameterize the waveform. 
Since the time parameterizations of the GWs and the orbital dynamics differ, we instead use the orbital frequency $\omegaorb$, which can be related to the wave frequency via Eq.~\eqref{eq:f22-forb}.

Following the discussion in Sec.~\ref{ssec:waveform_modes} about non-matching time parameterizations, we observe that under strong precession or residual eccentricity, $f^\mathrm{QA}_{22}$ may not be monotonic (see Fig.~\ref{fig:fdot} for reference).
To address this, alternative gauge-independent waveform-based measures have been proposed \cite{Boyle:2013nka, Varma:2019csw, Hamilton:2018fxk}.
In contrast, we use the orbit-averaged frequency $\langle f_{22}^\mathrm{QA} \rangle$, computed over two waveform cycles (one orbital period). This quantity offers a more robust alternative, ensuring monotonicity and preserving the approximate relation with the orbital frequency in Eq.~\eqref{eq:f22-forb}.
When the spin evolution of the merger waveform is not available, one can fall back to using reference frequency and spins in the metadata of the waveform.

\subsection{Fixing the frame axis}
\label{ssec:QL}

% QA frame(ref) /= LAL source-frame.
Once the spins and orbital frequency have been extracted from the NR simulation they must be transformed into the appropriate frame for generating the inspiral waveform, typically the \emph{LAL source-frame}. To achieve this, we use the precession angles at the reference frequency to construct the QA frame.
In this section we discuss the relation of the axis $\Qhat$ of the QA frame and of the axis of the \emph{LAL source-frame}, which is $\Lhat$, although in practice this is often replaced by the direction of the Newtonian angular momentum $\LNhat$, which is equivalent to the direction of the orbital angular frequency. In the next Sec.~\ref{ssec:orbital_plane} we then fix the orientation of the orbital plane so that the $\xhat$-axis points along $\nhat$.

% Q vs. L
To improve the approximation of $\Lhat$ beyond simply using~$\Qhat$, one could explore an iterative refinement procedure: generating inspiral waveforms with different choices of $\zhat(\tref)$ near to $\Qhat(\tref)$ and identifying the best choice by comparing the spin and $\Qhat$ evolutions with those of the merger waveform.
However, the spin evolution highly depends on the approximant used to generate the waveform. Therefore, they should only be trusted for small intervals after the reference point, and maybe this error dominates over the optimization one might pursue. We do not pursue this method further in this work.

Alternative options to using $\Qhat$ would be to compute either $\LNhat$ or a PN approximation for $\Lhat$ when information about the orbital evolution is known for the waveforms used. Here we choose to work with $\Qhat$ since it can be computed in a gauge-independent way and it requires no information beyond the waveform.
In Sec.~\ref{sec:VA} we study the difference between $\Lhat$ and $\Qhat$ across the parameter space and establish bounds on their deviation.

\subsection{Fixing the orbital plane rotation} \label{ssec:orbital_plane}

% ROTATE TO THE QA FRAME
We now have to describe the rotation of the orbital plane that fixes $\xhat = \nhat$. Since, $\alpha(t),\,\beta(t)$ completely determine $\xhat(t)$, this orbital plane fix can be determined with $\gamma(t)$. The minimal rotation condition, Eq.~\eqref{eq:MRC} only imposes a differential condition on $\gamma$, so there is freedom in redefining $\gamma(t) \to \gamma(t) + \Delta\gamma$.
Then, using Eq.~\eqref{eq:waveform_frames}, we can write
\begin{align}
    h^\mathrm{in}_{\ell m}(t) &= \sum_{m'=-\ell}^\ell h^\mathrm{QA}_{\ell m'}(t)\ \mathcal{D}^\ell_{m m'}\left(\alpha(t),\,\beta(t),\,\gamma(t)+\Delta\gamma\right) \quad \Longrightarrow \nonumber\\
    h^\mathrm{LAL}_{\ell m}(t) &= \sum_{m'=-\ell}^\ell h^\mathrm{in}_{\ell m'}(t)\ \mathcal{D}^{\ell}_{m m'}\left(-\gamma_\mathrm{ref}-\Delta\gamma,\,-\beta_\mathrm{ref},\,-\alpha_\mathrm{ref}\right) \label{eq:planefixing}
\end{align}

% FIXING THE ORBITAL PLANE IS NEEDED
In general, 
\begin{equation} \label{eq:n(x,y,phi)}
    \nhat(t) = \xhat(t) \cos{\phi(t)} + \yhat(t) \sin{\phi(t)}
\end{equation}
where $\phi(t)$ is the orbital phase, when the minimal rotation condition is met. In this framework,
\begin{equation} \label{eq:intomega}
    \phi=\int\omegaorb\ \mathrm{d}t.
\end{equation}
Equation~\eqref{eq:n(x,y,phi)} translates the $\nhat=\xhat$ identification at $t_{ref}$ to setting $\phi(t_{ref})=0$. This is possible by varying the integration constant in Eq.~\eqref{eq:intomega}, i.e. $\phi(t) \to \phi(t) + \Delta\phi$. Appendix~\ref{app:orbitalplane} justifies $\Delta\gamma = - \Delta\phi$.

% SOLVING IT
Should the merger description $\hA$ contain information about $\nhat(t)$ this degeneracy is broken by setting $\Delta\gamma = \phi(t_{ref})$.
However, if the description contains information of $\nhat$ at only one point, say $t_0$, we can use this fact to build
\begin{equation} \label{eq:4.6}
    \phi(t_1) = \phi(t_0) + \int_{t_0}^{t_1} \omegaorb(t)\  \mathrm{d}t.
\end{equation}
Usually, if the waveform comes from an NR simulation, the user can identify that in the initial point, $\xhat$ is either $\pm\nhat$.

% USING f22
Alternatively, employing Eq.~\eqref{eq:f22-forb}, one can write
\begin{equation} \label{eq:4.7}
    \phi(t_1) = \phi(t_0) + \pi \int_{t_0}^{t_1} f_{22}^\mathrm{QA}(t)\  \mathrm{d}t.
\end{equation}
Here, explicit orbit-averaging is unnecessary, as integration naturally smooths short-term variations if $t_1$ is sufficiently distant from $t_\mathrm{ref}$.
This approach is preferred when $\omegaorb$ is only sparsely sampled or not explicitly given in time. However, a subtle change in the time-parameterization has been introduced, for the expression shifted from purely dynamic-time parameterized in Eq.~\eqref{eq:4.6} to using the waveform time-parameterization in Eq.~\eqref{eq:4.7}. This relation holds under the assumption that the difference between these time parameterizations remains approximately constant throughout the inspiral regime.

If no information is known for $\nhat$, \cite{Estelles:2020twz} suggests $\Re{(h_{22})}>0,\, \Im{(h_{22})}=0,\Im{(h_{21})}<0$ is an equivalent criterion to $\xhat=\nhat$ and therefore can be used for identifying a time $t_0$ where $\phi(t_0)=0$, finally employing Eq.~\eqref{eq:4.7}.

\subsection{Hybridizing in the inertial frame} \label{ssec:hybrids_inertial}

% ALIGN DESCRIPTIONS
Once the two waveforms are generated, we need to align them in space and time. A natural idea is to split the alignment degrees of freedom into two categories: shifts and frame-fixing, as discussed in Sec.~\ref{ssec:dof}.
Aligning all the degrees of freedom at once has been performed in \cite{Sun:2024kmv}.
More specifically, 2 options can be considered: \begin{enumerate}
    \item Fix the same inertial frame for both descriptions, and then align the waveforms with the shifts.
    \item Use the simplicity of the non-precessing case by aligning the waveforms in the QA frame, and then fix the frames.
\end{enumerate}

% DISCUSS TWO OPTIONS
In the former approach, alignment is performed in an inertial frame, which is expected to be noisier than aligning non-precessing waveforms. Choosing an appropriate inertial frame is therefore essential.
In contrast, the latter approach retains the simplicity of aligning non-precessing waveforms but requires the delicate task of matching the time-dependent coprecessing frames through an inertial rotation. We have already expressed that the latter is preferred due to the method's robustness. Despite that, in this section we sketch an implementation of inertial-frame alignment, that is not part of the final algorithm.

% WHAT TO DO
A few previous works have explored inertial-frame hybridization. In Ref.~\cite{Klein:2015hvg}, PN-NR TD hybrid waveforms are constructed by minimizing the phase difference of $\Psi_4$ over a window.
Ref.~\cite{Sadiq:2020hti} defines discrepancy as the integrated square difference summed across all available modes, and uses waveform approximants other than PN expansions, applying it to mildly precessing systems. Broader applicability remains unexplored.

% DESCRIBING INERTIAL FRAMES
The first step is to express both waveforms in a common inertial frame.
While waveforms may be provided in arbitrary frames, most models adopt the \emph{LAL source-frame} at a specific reference point.
This frame can be (approximately) identified in a gauge-independent manner using the coprecessing angles at the reference point, as described in Eq.~\eqref{eq:planefixing}.
We choose to use the LAL source-frame for alignment, due to its practical convenience and resemblance to the coprecessing frame at the vicinity of the reference point, which minimizes the amplitudes of non-dominant $\ell=2$ modes, in favour of the dominant modes.
Once the QA axis is aligned, the orbital plane orientation can be adjusted via a simple shift in $\varphi_0$. 

% HYBRIDIZATION INTERVAL
To leverage these properties, we center the alignment interval on the reference point. Earlier intervals retain more of the original merger waveform (which is the most accurate waveform in many applications) but may suffer from residual eccentricity; while later intervals benefit from circularization. NR waveforms may be affected by junk radiation or other artifacts at the beginning of the waveform.
The interval must be long enough to smooth noise and eccentricity-induced oscillations; single-point alignments are generally unreliable and thus avoided.
Here, we adopt a symmetric interval around the reference point, with a fixed length.
Optimization of the interval length is possible
out optimizing its length case by case; the sensitivity to this choice was studied throughout the development of the method with no visible trend.

% DISCREPANCY MEASURE
The discrepancy $\delta$ between waveforms is minimized over time, phase and polarization shifts and defined as
\begin{equation} \label{eq:alignment_inertial}
    \delta = \sum_{l,m\,\in\text{ modes}} c_{lm} \int \left| h^A(t) - h^B(t-t_0) \cdot \mathrm{e}^{i(m\varphi_0+2\psi_0)} \right|\ \mathrm{d}t
\end{equation}
as in \texttt{NRHybSur3dq8}~\cite{Varma:2018mmi}.
We typically choose $c_{22}=c_{21}=1$ and zero otherwise, to emphasize alignment of the $(2,2)$ and $(2,1)$ modes.
Mode mixing is limited near the reference point due to the use of the LAL source-frame. 
Despite that, it still complicates direct alignment due to the impracticality of decomposing modes into amplitudes and frequencies as we will do for the coprecessing-frame alignment in the next Sec.~\ref{ssec:hybrids_coprecessing}.

The optimal time shift $t_0$ is usually close to $t_{\mathrm{peak},A} - t_{\mathrm{peak},B}$ within a few M. The polarization angle $\psi_0$ can be assessed independently by checking the tetrad conventions of both waveforms.

% NUMERICAL OPTIMIZATION
Despite its formal similarity to non-precessing multimode alignment, inertial-frame alignment features a more irregular discrepancy function.
Short windows often lead to overfitting, with false global minima driving $t_0$ away from physically reasonable values, particularly in strongly precessing or high-mass ratio systems, and at late times where models diverge more.
In such cases, estimates of $t_0$ based on peak times help determine the reliability of the optimization and eventually motivate the choice of a different local minima.
Illustrative failures of inertial-frame alignment are presented in Sec.~\ref{ssec:false_minima}.
These issues support our preference for QA frame alignment as described in Sec.~\ref{ssec:hybrids_coprecessing}.

\subsection{Hybridizing in the coprecessing frame} \label{ssec:hybrids_coprecessing}

% MOTIVATION
Alignment in the QA frame for building %$\Psi_4$
PN-NR hybrids was explored in \cite{Schmidt:2012rh} for the dominant (2,2) mode only. In this work we perform multi-mode alignment.

% HYBRID INTERVAL
We define a hybridization interval centered on the reference point and extending a few orbits (usually $N=8$). We follow this simple choice to ensure that the physical parameters of both waveforms, that have been matched at the reference point, have not yet drifted away by different evolutions from different models.

% HYERARCHICAL ALIGNMENT
To determine the best shift values, we follow a hierarchical strategy: we first compare the frequencies of the 22 mode in the QA frame to determine the time shift $t_0$. This typically yields $t_0 \sim t_{\mathrm{peak},A} - t_{\mathrm{peak},B}$. Although mode amplitudes could also be used to determine $t_0$ their slow variation ($\delta|h|/|h|\ll1$) makes the result sensitive to even tiny power losses in the NR waveforms.
% PHASE SHIFTS
The phase and polarization shift values $\varphi_0$ and $\psi_0$ are then found by minimizing the phase difference across modes within the hybridization interval:
\begin{equation}
    \delta = \sum_{\ell, m} c_{\ell, m} \int_{t_1}^{t_2} \left[ \varphi^A_{\ell, m} - \left( \varphi^B_{\ell, m} + m\varphi_0 + 2 \psi_0 \right) \right]
\end{equation}
where $c_{\ell m} = \max_{t}{|h_{\ell m}(t)|}$ weights dominant modes.

Care is taken to avoid configurations where $\beta=0$ inside the hybridization interval, since that causes $\alpha$ to jump by $\pi\ \mathrm{rad}$ and lead to ill-defined QA angles. To mitigate this, we apply an artificial inertial-frame rotation to both waveforms. If the artificial rotation is chosen properly, this causes both frames to rotate synchronously to a configuration where $\beta$ is far from the value of zero during the whole evolution of the system.
The quaternion description of rotations would provide an alternative solution to this issue, and we leave this for future work.

\begin{figure*}
    \includegraphics[width=0.31\linewidth]{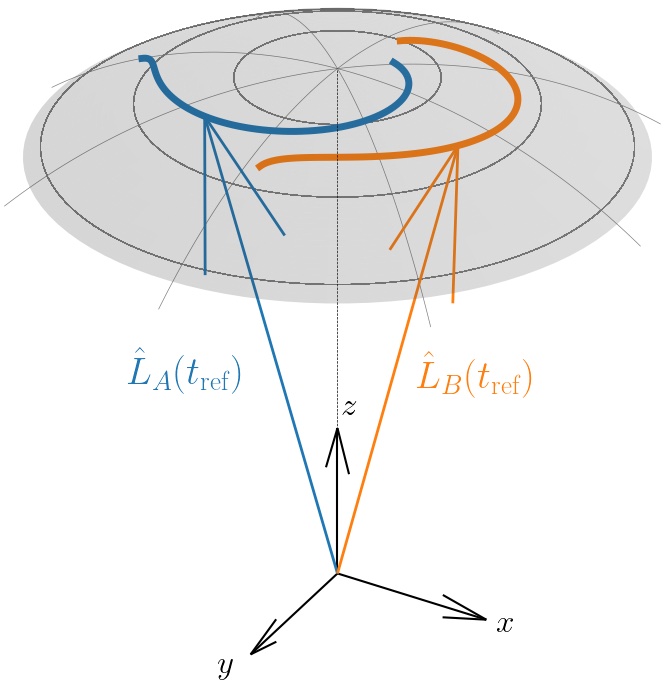}
    \includegraphics[width=0.33\linewidth]{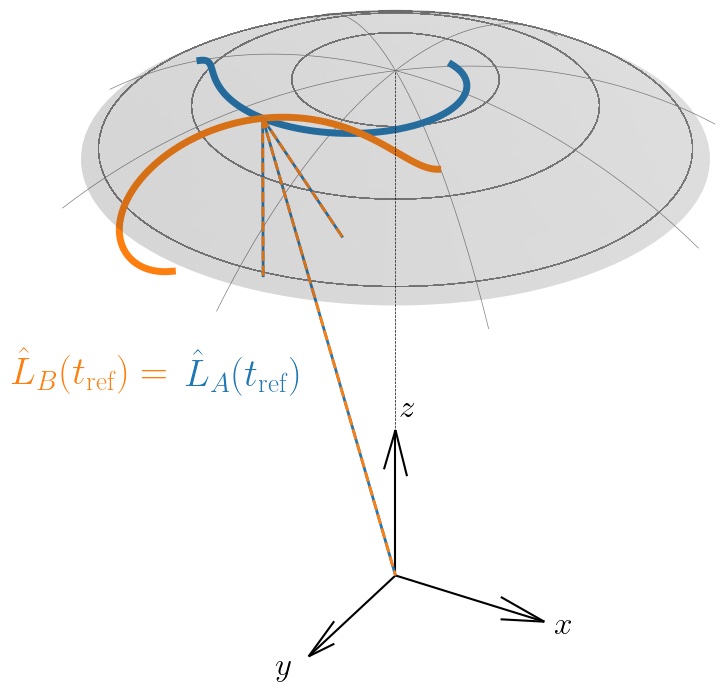}
    \includegraphics[width=0.33\linewidth]{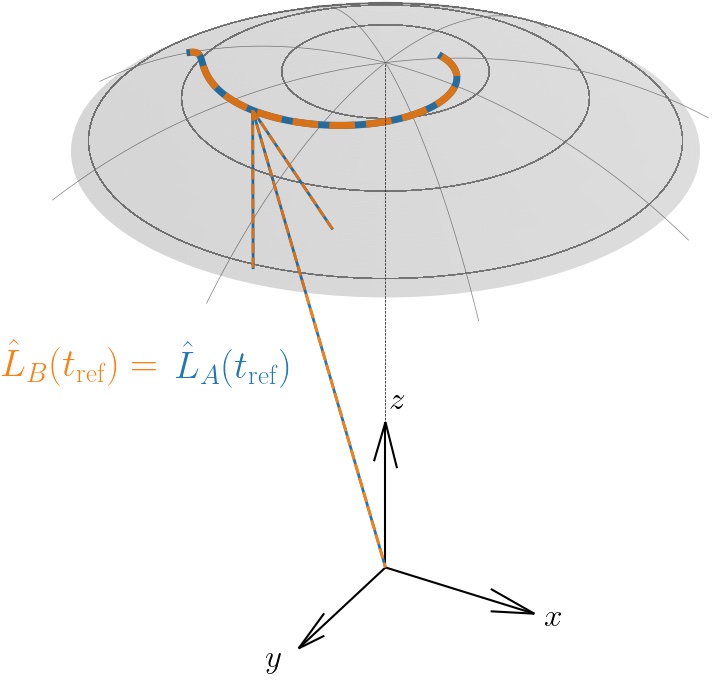}
    \caption{Illustration of the frame-fixing: orientation of the $\zhat$-axis of the QA frame for two descriptions (A and B) of the same system, with respect to an inertial frame. In the left panel, we visualize the trajectory of $\Lhat(t)$ on the unit sphere for two descriptions (in two different inertial frames) of the same system. One can acknowledge the similar shape of the two trajectories. The center panel shows an unsatisfactory choice of the freedom, where two degrees of freedom have been used to match the trajectories at a single point, like the first step in the procedure in \cite{Schmidt:2012rh}. The right panel shows the right choice of fixing the freedom, which uses three degrees of freedom and for the whole trajectory is aligned.}
    \label{fig:frame_fixing}
\end{figure*}

% FRAME FIXING
With the QA frame waveforms aligned, we next align the precession angles. Each description provides time-dependent angles $\{\alpha(t),\,\beta(t),\,\gamma(t)\}$ and corresponding rotation matrices $\bm{R}(t)$. The rotation between frames is then
\begin{equation} \label{eq:R0}
    \bm{R}_0(t) = \bm{R}_\mathrm{B}(t) \cdot {\bm{R}_\mathrm{A}(t)}^\mathrm{T},
\end{equation}
which transforms vectors in frame A to frame B. While this expression appears time-dependent, in the proxy case where waveforms A and B represent the same physical system and are only expressed in different inertial frames, $\mathbf{R}_0$ is exactly time-independent, up to numerical precision. This is because, in that case, the time-dependent QA frame of waveforms A and B is exactly the same. Therefore, $\bm{R}_0$ can be parameterized by three Euler angles ${\alpha_0,\,\beta_0,\,\gamma_0}$.
This observation is the key step of the precession angles alignment.

Nevertheless, in the general case where the waveforms are generated with different models and correspond to physical systems only similar to each other, gauge inconsistencies and different binary evolution from different models will make the rotation $\mathbf{R_0}$ only approximately inertial, with the approximation being more exact near the point where the second description was generated.
Then, to find $\mathbf{R}_0$, a naive strategy would be to evaluate it at a single reference time $t_\mathrm{ref}$, but this requires perfect orbital plane fixing and is sensitive to local noise. We instead minimize a discrepancy over the hybrid interval:
\begin{equation} \label{eq:criteria_precessing}
    \delta = \int_{t_1}^{t_2} \left( \left| \mathbf{R}_0 \cdot \hat{L}_A(t) - \hat{L}_B(t) \right| + \left| \mathbf{R}_0\cdot \hat{n}_A(t) - \hat{n}_B(t) \right| \right)\ \mathrm{d}t.
\end{equation}

This frame-fixing procedure generalizes the method of \cite{Schmidt:2012rh}, which aligns $\Qhat$ at a single point. Since this leaves one rotational degree of freedom unconstrained, they additionally match trajectory curvature. Then,
\begin{equation}
    \mathbf{R}'_\mathrm{A}(t) = \mathbf{R}_0 \cdot \mathbf{R}_\mathrm{A}(t),
\end{equation}
where the optimization of $\alpha_0,\,\beta_0,\,\gamma_0$ in $\mathbf{R}_0$ reduces the difference between $\mathbf{R}'_\mathrm{A}(t)$ and  $\mathbf{R}_\mathrm{B}(t)$.

% FIGURE
Figure~\ref{fig:frame_fixing} illustrates the frame-fixing procedure. The left panel shows the two unaligned trajectories, while the center panel shows a rotated version of one of the trajectories, matching the other trajectory at a single point. 
Trajectories are then completely aligned at the right panel. Since the trajectories coincide exactly after alignment, this figure represents a proxy case where the waveform models and the intrinsic parameters are identical. In real cases, alignment is only approximate and maximized for the portion of the trajectory within the hybridization interval. 

A key complication arises from the freedom in the choice of $\gamma$ in each description. Even in the proxy case, inconsistent choices for $\gamma$ make $\mathbf{R}_0$ time-dependent.
Geometrically, aligning $\nhat$ with $\xhat$ for both waveforms requires consistent definitions of $\gamma$.
More generally, the condition can be relaxed and the angle between $\nhat$ and $\xhat$ must match across the two waveforms to allow for a time-independent $\mathbf{R}_0$.
This can be understood through the explicit rotation decomposition (Eq.~\eqref{eq:rotation_decomposition}) as
\begin{equation} \label{eq:R0_as_composition}
\begin{split}
    \bm{R}_0(t) &= \bm{R}_\mathrm{B}(t) \cdot {\left[ \bm{R}_\mathrm{A}(t) \right]}^\mathrm{T} = \\
    &= \mathcal{R}_z(\alphaB) \ \mathcal{R}_y(\betaB) \ \mathcal{R}_z(\gammaB) \ \mathcal{R}_z(-\gammaA) \ \mathcal{R}_y(-\betaA) \ \mathcal{R}_z(-\alphaA) = \\
    &= \mathcal{R}_z(\alphaB) \ \mathcal{R}_y(\betaB) \ \mathcal{R}_z(\gammaB-\gammaA) \ \mathcal{R}_y(-\betaA) \ \mathcal{R}_z(-\alphaA)
\end{split}
\end{equation}
confirming that only the relative offset $\gamma_B - \gamma_A$ affects $\mathbf{R}_0$.

Although orbital plane fixing described in Sec.~\ref{ssec:orbital_plane} can enforce consistent $\gamma$ values, we adopt a more flexible approach by determining only the relative shift $\Delta\gamma$ that minimizes the time variation of $\mathbf{R}_0$ over the hybridization interval.
Specifically, we minimize the variance of the $r_{zz}$ component:
\begin{equation}
    \Delta\gamma = \arg\min, \mathrm{var}{(t_1,t_2)}(r_{zz}),
\end{equation}
where
\begin{equation}
    r_{zz}(t) = \sin\beta_A \sin\beta_B \cos(\gamma_B - \gamma_A + \Delta\gamma) + \cos\beta_A \cos\beta_B,
\end{equation}
and the variance is defined as
\begin{align}
    \mathrm{var}{(t_1,t_2)}(f) &= \frac{1}{t_2 - t_1} \int_{t_1}^{t_2} \left(f(t) - \langle f \rangle \right)^2\ \mathrm{d}t, \\
    \langle f \rangle &= \frac{1}{t_2 - t_1} \int_{t_1}^{t_2} f(t)\ \mathrm{d}t.
\end{align}

Once $\Delta\gamma$ is determined, it is applied to $\gamma_\mathrm{B}$, and $\mathbf{R}_0$ is computed. Full expressions for the components of $\mathbf{R}_0$ in terms of the precession angles are given in App.~\ref{app:composition}.

\subsection{Blending functions} \label{ssec:blending}

% BLENDING FUNCTION
Once the waveforms and the precessing angles are aligned they can be combined over a hybridization interval using a blending function like Eq.~\eqref{eq:blending}.
In our method, the hybridization interval $(t_1,t_2)$ is chosen to coincide with the alignment window, ensuring that the discrepancy between the input waveforms is minimized where they are blended, thus yielding maximal agreement. Each mode's amplitude and phase is blended separately using the window function
\begin{equation} \label{eq:blending}
    w(t) = \dfrac{1}{2}\ \left[1 + \cos{\left(\dfrac{\pi(t-t_1)}{t_2-t_1} \right)} \right],
\end{equation} which smoothly interpolates between the two waveforms in Eq.~\eqref{eq:blend}.
The precession angles are also hybridized using the same function. The resulting 
QA frame waveform is then rotated back to an inertial frame using the hybridized Euler angles.

\section{Results} \label{sec:discussion}
In this section, we present the performance of our method with some illustrative examples supporting the applicability of the method and highlighting some features that were discussed in the previous Section.
We use NR simulations from the \texttt{sxs}~catalog~\cite{Boyle:2019kee} as well as the \texttt{lalsimulation}~\cite{lalsuite} implementation of \texttt{IMRPhenomTPHM}~\cite{Estelles:2021gvs} and \texttt{SEOBNRv5PHM}~\cite{Ramos-Buades:2023ehm} using the \texttt{pyseobnr}~module~\cite{Mihaylov:2023bkc}.

\begin{table*}
\centering
\begin{tabular}{p{1.6cm} p{2.5cm} p{2.5cm} p{2.5cm} p{3.0cm}}
\hline
    \textbf{Parameter} & System A & System B & System C & System D \\
    & Non-precessing & Mild precession & Strong precession & GW200129-like~\cite{Varma:2022pld} \\
\hline \hline
    $q$ & $2$ & $2$ & $6$ & $2.1$ \\ \hline
    $\chi_\mathrm{1,ref}$  & $(0,\,0,\,0.5)$   & $(0.6,\,0,\,0.5)$   & $(0.7,\,0.3,\,-0.5)$& $(-0.46,\,0.80,\,0.21)$ \\ \hline
    $\chi_\mathrm{2,ref}$  & $(0,\,0,\,0.5)$   & $(0,\,0.3,\,0.5)$   & $(-0.2,\,0,\,-0.25)$& $(0.01,\, -0.41,\,-0.55)$ \\ \hline
    $\omegaref$  & $0.01$ & $0.01$ & $0.012$ & $0.00734$ \\ \hline
\hline
\end{tabular}
\caption{Parameters of the simulations used in the results discussion in Sec.~\ref{sec:discussion}.}
\label{tab:discussion_parameters}
\end{table*}

% GENERATING THE INSPIRAL WAVEFORM
\subsection{Generating the inspiral waveform} \label{sec:VA}

% CHALLENGES
Generating an inspiral waveform with the same physical parameters as the merger waveform is crucial, as shown in steps 2 and 3 of our algorithm and discussed in Secs.~\ref{ssec:QL}~\&~\ref{ssec:orbital_plane}. We identify three main reasons why the waveforms might not coincide:
\begin{enumerate}
    \item The QA frame only approximates (but is not identical to) the \emph{LAL source frame}, as discussed in Sec.~\ref{ssec:QL}. Therefore, the spins of both waveforms are slightly different, even when both waveforms use the same model. This effect is discussed in this Section.
    \item The approximant used for the inspiral waveform is in general different than the method used to generate the merger waveform (either a waveform model or an NR product). Therefore, we expect the spins to be subject to gauge discrepancies. As mentioned before, in this work we do not discuss the effect of gauge discrepancies.
    \item Even with optimized intrinsic parameters, the two waveforms will not match perfectly due to systematic differences within the models. Refer to Sec.~\ref{ssec:systematics} for a discussion.
\end{enumerate}

%%%%% FIRST CHALLENGE %%%%%
\begin{figure}
    \includegraphics[width=\columnwidth]{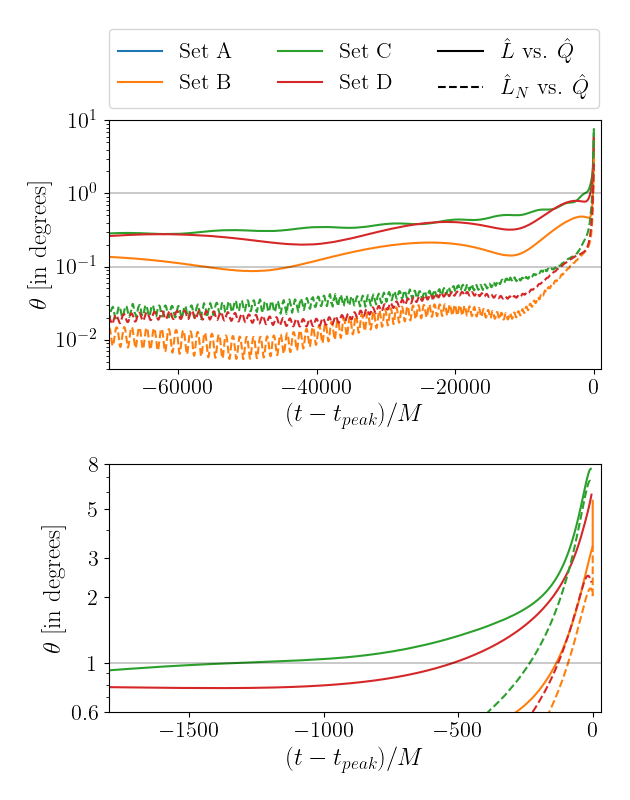}
    \caption{Comparison of $\Lhat$, $\Lhat_N$ and $\Qhat$ for the 4 waveforms described in Tab.~\ref{tab:discussion_parameters} generated with the \texttt{SEOBNRv5PHM} approximant. 
    Colors correspond to Systems B (orange), C (green), and D (red), comparing $\Qhat$ with $\Lhat$ (solid lines) and $\Lhat_N$ (dashed lines).
    The line corresponding to System A (non-precessing waveform) is exactly $0$ since, by construction, $\Lhat=\Lhat_N=\Qhat=\zhat$. As expected, precessing effects become more important later in the evolution, and with them, the difference between $\Qhat$ and $\Lhat$ increases. The lower panel shows a more detailed view of the last part of the evolution.}
    \label{fig:discussion1a}
\end{figure}

\begin{figure}
    \includegraphics[width=\columnwidth]{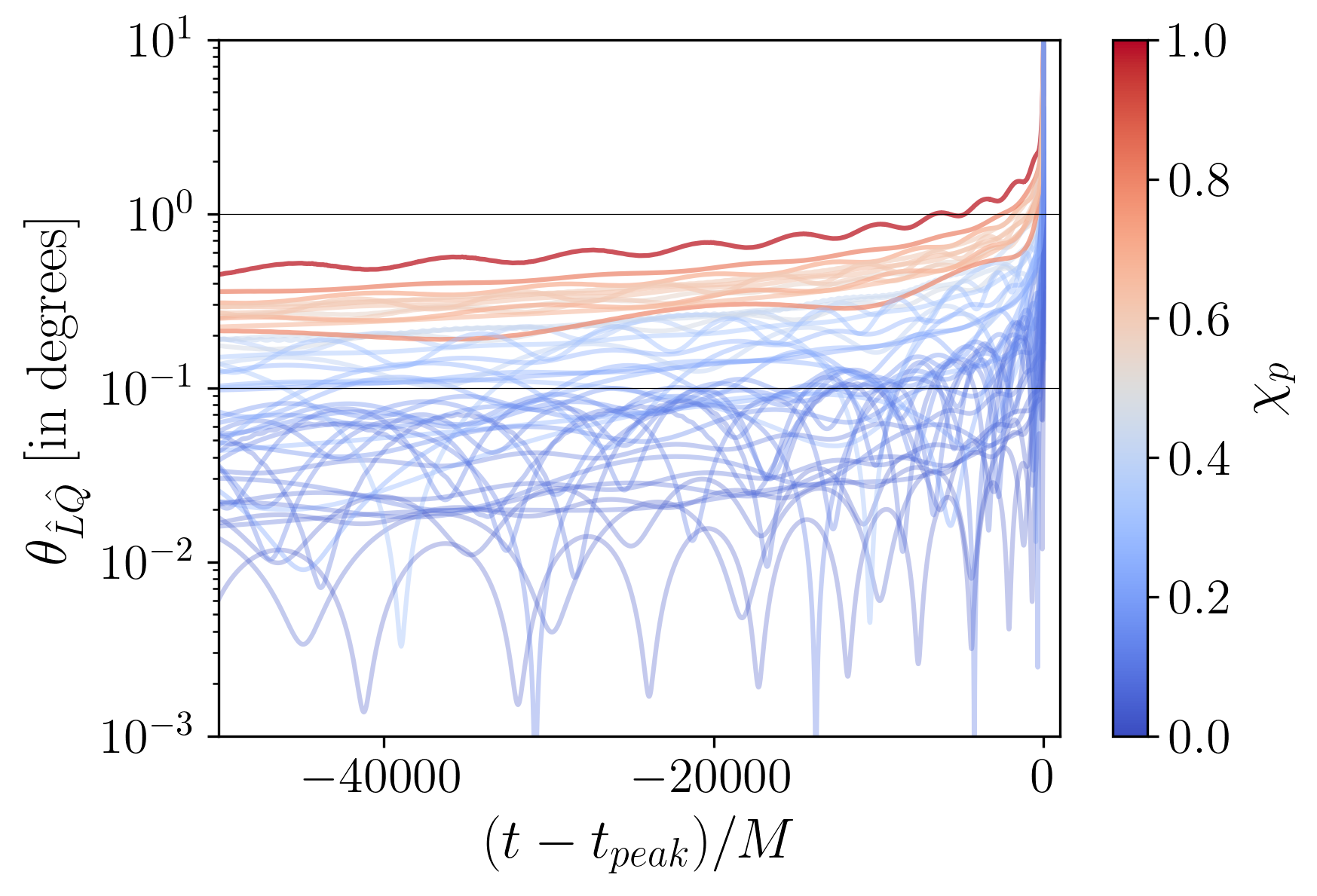}
    \includegraphics[width=\columnwidth]{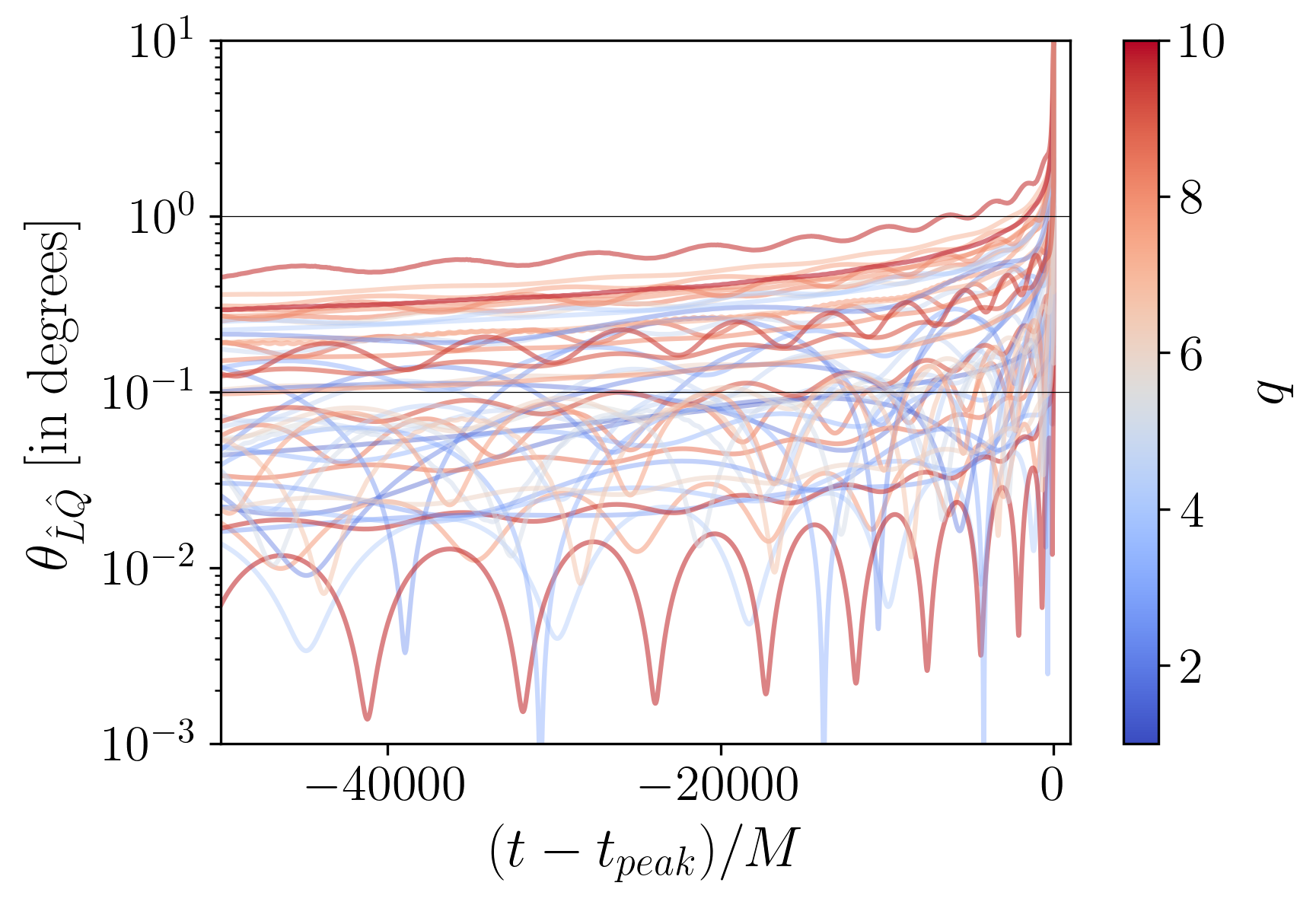}
    \caption{Comparison of $\Lhat$ and $\Qhat$ for a set of 50 waveforms generated with \texttt{SEOBNRv5PHM} approximant. Each waveform is represented by a line and its color corresponds to its effective precessing spin $\chi_p$ (top panel) or mass ratio $q$ (bottom panel). Systems with higher $\chi_p$ exhibit larger differences between $\Qhat$ and $\Lhat$, with all values remaining below $1^\circ$ until the last few orbits.}
    \label{fig:discussion1b}
\end{figure}
\begin{figure}
    \includegraphics[width=\columnwidth]{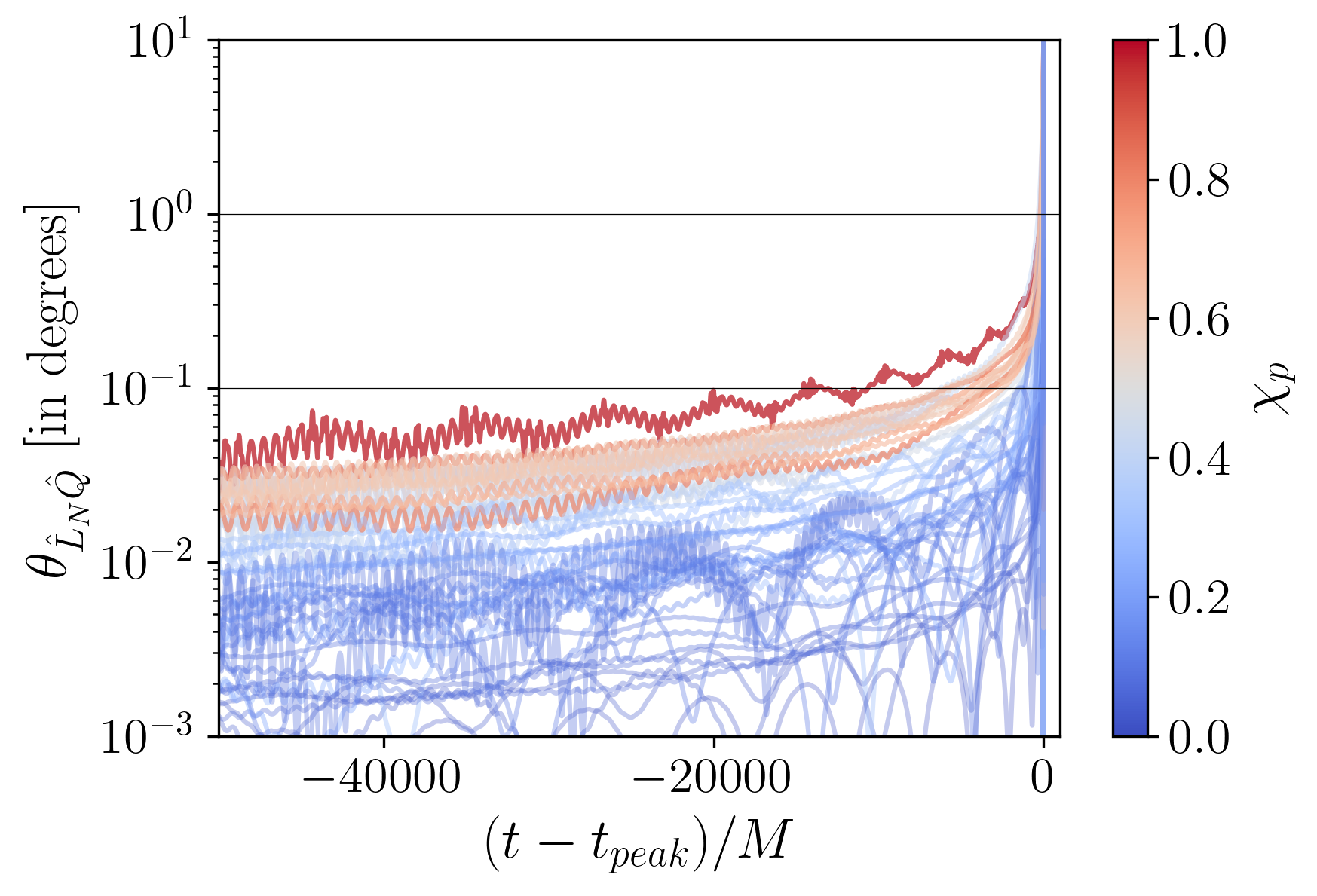}
    \includegraphics[width=\columnwidth]{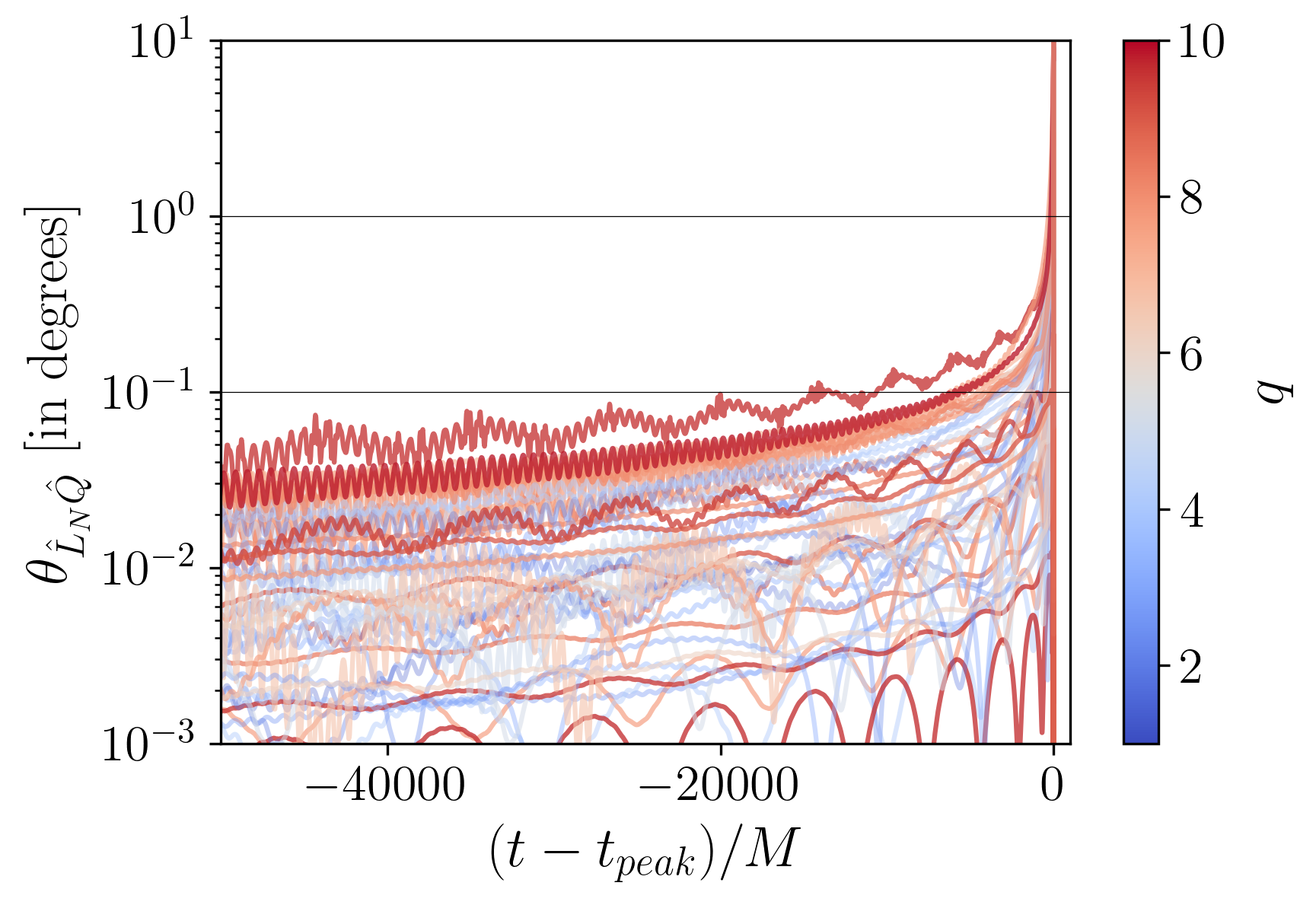}
    \caption{Comparison of $\LNhat$ and $\Qhat$ for the same set of waveforms as in Fig.~\ref{fig:discussion1b}. Each line represents a waveform, colored by effective precessing spin $\chi_p$ (top panel) or mass ratio $q$ (bottom panel). Systems with higher $\chi_p$ show larger differences between $\Qhat$ and $\Lhat$. Relative to Fig.~\ref{fig:discussion1b}, we observe that $\LNhat$ more closely tracks $\Qhat$ than $\Lhat$ does. The nutation present in $\LNhat$ manifests as high-frequency oscillations in this Figure.}
    \label{fig:discussion1c}
\end{figure}

We now compare the direction of maximum emission $\Qhat$ computed from the waveform to the direction of $\Lhat$ provided by the model.
% 4 SYSTEMS
For the 4 systems in Tab.~\ref{tab:discussion_parameters}, we generate a waveform with the \texttt{SEOBNRv5PHM} model and compute the angle between $\Lhat-\Qhat$, and we observe the angle is $\lesssim 1^\circ$ up to the last $\sim10^3\ \mathrm{M}$ of evolution, as seen in Fig.~\ref{fig:discussion1a}.

% SEOBNRv5PHM SETS
A further test with a set of 50 waveforms generated with \texttt{SEOBNRv5PHM} produces similar results. This set was generated with $q$ uniformly sampled in $[1,10)$, $|\chi_i|$ uniformly sampled in $[0,1)$, and $\hat{\chi_i}$ isotropically sampled. Then, differences between $\Qhat$ and both $\Lhat$ and $\LNhat = \rhat \times \dot{\rhat}$ are computed.
In Figs.~\ref{fig:discussion1b}~and~\ref{fig:discussion1c} we explore the dependence of this result with $q$ and $\chi_p = \max{\left( \chi_{1\perp},\, \dfrac{4+3q}{4q^2+3q}\ \chi_{2\perp} \right)}$. We find that both higher $q$ and $\chi_p$ increase the discrepancy between the maximum emission direction and the orbital angular momentum. We also observe how $\LNhat$ shows nutation while $\Lhat$ does not, as expected \cite{Boyle:2014ioa}.

The difference between $\Qhat$ and $\Lhat$ can complicate alignment by making the inspiral and merger waveforms describe different systems. This can be resolved if the merger waveform provides the evolution of $\Lhat$ (and $\nhat$) during the inspiral. As noted in Sec.~\ref{ssec:QL}, $\LNhat$ or a PN approximant to 
$\Lhat$ can be used when orbit evolution is available, or alternatively, an iterative procedure can refine the $\Qhat$ approximation of $\Lhat$.
However, we choose to describe our method using $\Qhat$ since it can be extracted directly from the waveform and we find the approximation satisfactory.

\subsection{Waveform systematics} \label{ssec:systematics}

An additional source of discrepancy arises from waveform model systematics and their disagreements with NR simulations, together with the misalignment of gauge conditions across models.

% QUANTITIES TO DESCRIBE
To test these differences, we explore the waveform systematics for the 4 systems in Tab.~\ref{tab:discussion_parameters}.
A handful of metrics can be used to indicate the similarity of two waveforms. The most commonly used is the mismatch $\mathcal{MM}$,
\begin{equation}
    \mathcal{MM}\left( h^\mathrm{A},h^\mathrm{B} \right) = \max_{t_0,\varphi_0} \dfrac{ \langle h_1, h_2 \rangle}{\sqrt{\langle h_1, h_1 \rangle \cdot \langle h_2, h_2 \rangle}},
\end{equation}
where $t_0,\,\varphi_0$ are the time and phase shifts already introduced in the previous sections, and the inner product in the space of the waveforms is defined as
\begin{equation}
    \langle h_1, h_2 \rangle = 4\ \Re \int_{f_\mathrm{min}}^{f_\mathrm{max}} \dfrac{ \tilde{h}_1(f)\ \tilde{h}_2^\ast(f)}{S_n(f)},
\end{equation}
with $(f_{min},f_{max}) = (20,1024)\ \mathrm{Hz}$ and $S_n(f)$ follows the LIGO A+ noise curve \cite{LIGO-T1800042}.

Other quantities can be used to supplement the comparison. For instance, regarding the amplitude, the \emph{relative amplitude difference} is measured as
\begin{equation}
    \delta|h_{\ell m}| = \dfrac{|h_{\ell m}^\mathrm{A}|-|h_{\ell m}^\mathrm{B}|}{|h_{\ell m}^\mathrm{A}|}
\end{equation}
for a given mode $(\ell,m)$ and frame.
In the inertial frame, the amplitudes of subdominant modes exhibit strong precession-induced modulations; hence, we restrict this measure to the $(2,2)$ mode.

Similarly, the \emph{phase difference} is defined as
\begin{equation}
    \Delta\varphi_{\ell m}(t) = {\varphi_{\ell m}}_\mathrm{A}(t) - {\varphi_{\ell m}}_\mathrm{B}(t)
\end{equation}
for a given mode $(\ell,m)$ and frame.
Since the waveforms can be reparameterized with a choice of $\varphi_0$, artificially affecting the measure, we set ${\varphi_{\ell m}}(t=-1000\ \mathrm{M})=0$ for both waveforms. In this way, when $t\to\infty$ we recover the accumulated phase difference of the two waveforms up to the last $1000\ \mathrm{M}$ before coalescence.
In the inertial frame, the phase of the higher-order modes is dominated by that of the dominant modes, making an accurate interpretation of the phase difference unreliable; for this reason, we do not compute it.

Moreover, differences between models are expected to increase closer to merger but we are not interested in comparing them at merger but rather before the last few cycles. For this reason, we cut the computation at $t-t_{peak}<-1000\ \mathrm{M}$ and we use the maximum relative amplitude and the median in the last $10000\ \mathrm{M}$ before that mark.

% SEOB vs. TPHM
Table~\ref{tab:comparison_models} shows the maximum of $\Delta\varphi_{\ell m}$ and $\delta|h_{\ell m}|$ with the median value of $\delta|h_{\ell m}|$.
Such metrics in the inertial frame are only shown for the dominant modes, since subdominant modes amplitude is strongly modulated by precession and its phase is dominated by the (2,|2|)-mode phase.

\begin{table}
\centering
\begin{tabular}{m{1.8cm}m{1.0cm}m{1.2cm}m{1.2cm}p{1.2cm}m{1.2cm}}
\hline
    \textbf{Comparison} & Frame & System A & System B & System C & System D \\
\hline \hline
    \multirow{2}{*}{$\max\Delta\varphi_{22}$} 
    & Inertial  & \multirow{2}{*}{$9\cdot10^{-2}$}  & $9\cdot10^{-1}$   & -     & $1.2$ \\
    & QA        &                                   & $9\cdot10^{-1}$   & $3.2$   & $1.4$ \\ \hline
    \multirow{2}{*}{$\max\delta|h_{22}|$} 
    & Inertial  & \multirow{2}{*}{$2\cdot10^{-3}$}  & $4\cdot10^{-2}$   & -     & $0.1$ \\
    & QA        &                                   & $4\cdot10^{-3}$   & $3\cdot10^{-3}$   & $3\cdot10^{-3}$ \\ \hline
    \multirow{2}{*}{$\mathrm{med}\ \delta|h_{22}|$}
    & Inertial  & \multirow{2}{*}{$3\cdot10^{-4}$}  & $1\cdot10^{-2}$   & $2\cdot10^{-1}$   & $2\cdot10^{-2}$ \\
    & QA        &                                   & $5\cdot10^{-4}$   & $1\cdot10^{-3}$   & $8\cdot10^{-4}$ \\ \hline
    %\multirow{2}{*}{$\mathcal{MM}_{22}$}
    %& Inertial  & \multirow{2}{*}{$3\cdot10^{-4}$}  & $1\cdot10^{-2}$   & $2\cdot10^{-1}$   & $2\cdot10^{-2}$ \\
    %& QA        &                                   & $5\cdot10^{-4}$   & $1\cdot10^{-3}$   & $8\cdot10^{-4}$ \\ \hline
\hline
    $\max\Delta\varphi_{21}$        & QA    & $5\cdot10^{-2}$   & $4\cdot10^{-1}$   & $1.6$   & $0.7$ \\ \hline
    $\max\delta|h_{21}|$            & QA    & $2\cdot10^{-3}$   & $9\cdot10^{-2}$   & $3\cdot10^{-2}$   & $4\cdot10^{-1}$ \\ \hline
    $\mathrm{med}\ \delta|h_{21}|$  & QA    & $7\cdot10^{-4}$   & $4\cdot10^{-2}$   & $1\cdot10^{-2}$   & $2\cdot10^{-1}$ \\ \hline
    %$\mathcal{MM}_{22}$             & QA    & $7\cdot10^{-4}$   & $4\cdot10^{-2}$   & $1\cdot10^{-2}$   & $2\cdot10^{-1}$ \\ \hline
\hline
\end{tabular}
\caption{Discrepancy measures for the 4 systems in Table~\ref{tab:discussion_parameters}. For each system, waveforms are generated with both \texttt{IMRPhenomTPHM} and \texttt{SEOBNRv5PHM}, only considered up to 1000 M before merger. The table shows the maximum relative amplitude difference and the maximum phase difference accumulated in that region, as well as the median relative amplitude difference in the 10000 M before the cut. Phase differences are shown in radians. As expected, differences are larger for systems with higher precession.}
\label{tab:comparison_models}
\end{table}
\begin{table}
\centering
\begin{tabular}{m{1.2cm}m{1.2cm}m{1.2cm}m{1.2cm}p{1.2cm}m{1.2cm}}
\hline
    Mode & Frame & System A & System B & System C & System D \\
\hline \hline
    \multirow{2}{*}{(2,2)} & Inertial  & \multirow{2}{*}{$7\cdot10^{-5}$}  & $5\cdot10^{-3}$   & $2\cdot10^{-1}$ & $4\cdot10^{-1}$ \\
          & QA        &                                   & $8\cdot10^{-5}$   & $7\cdot10^{-3}$   & $6\cdot10^{-4}$ \\ \hline
    (2,1) & QA        &  $2\cdot10^{-3}$   & $2\cdot10^{-2}$   & $8\cdot10^{-1}$ & $8\cdot10^{-2}$ \\ \hline
\hline
\end{tabular}
\caption{Mismatches of waveforms generated with \texttt{SEOBNRv5PHM} and \texttt{IMRPhenomTPHM} models for the configurations in Table~\ref{tab:discussion_parameters} and $M=60\ \Msun$. As expected, differences are larger for the subdominant mode (2,1) and for inertial frame modes.}
\label{tab:mismatches_models}
\end{table}

We observe that $\Delta\varphi_{22}$ can reach values of about $\sim1\ \mathrm{rad}$ before the last $1000\ \mathrm{M}$, even in the QA frame, with the (2,1) mode phase difference being half as large. Regarding the amplitude difference, it should stay in $\sim \mathrm{few}\ 10^{-3}$ in the QA frame, while in the inertial frame high precessing systems exhibit higher discrepancies. The (2,1) mode can have much higher differences.

Table~\ref{tab:mismatches_models} shows the mismatches obtained when comparing the \texttt{SEOBNRv5PHM} and the \texttt{IMRPhenomTPHM} waveforms for a total mass of $M=60\ \mathrm{M}_\odot$. We see how mismatches are progressively worse for systems with more precession.

% SXS
For another test of waveform systematics with NR waveforms, hybrids are built with the 29 simulations of the SXS Collaboration's second catalog of BBH simulations \cite{Boyle:2019kee} that exhibit precession and are considerably long (initial separation $\geq 20\ \mathrm{M}$). For all these simulations, the initial frequency is $\lesssim 30\ \mathrm{Hz}$ for $M=20\ \mathrm{M}_\odot$. These waveforms are hybridized with a \texttt{IMRPhenomTPHM} waveform with initial frequency $20\ \mathrm{Hz}$ and hybridized at $f_{22ref} = 40\ \mathrm{Hz}$.
This means 72-95\% of the hybrid waveform was originally from the model waveform and the rest from the NR waveform.

\begin{figure*}
    \includegraphics[width=\linewidth]{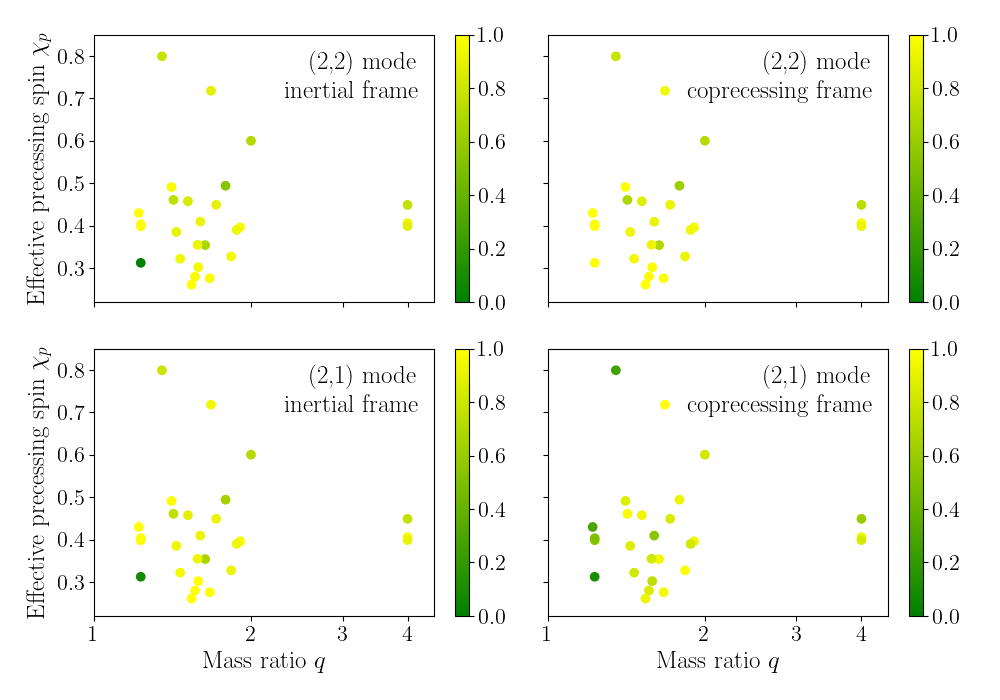}
    \caption{Mismatch ratio $\mathcal{MM}_\mathrm{NR-hybrid} / \mathcal{MM}_\mathrm{NR-TPHM}$ for (2,2) and (2,1) modes in an inertial and coprecessing frame of 29 precessing long SXS simulations for $M=20\ \Msun$. Horizontal axis indicates mass ratio $q$, vertical axis indicates $\chi_p$ as defined earlier in this section.
    Smaller values of the ratio indicate the Hybrid is ``closer'' to the NR simulation than the TPHM waveform is. Values much larger than 1 would hint possible waveform artifacts introduced during the hybridization process. The maximum of all ratios in this figure is $\leq 1.008$.}
    \label{fig:mismatches1}
\end{figure*}

In Figure~\ref{fig:mismatches1} we show the mismatch comparisons for this set of simulations, indicating their mass ratio and effective precessing spin $\chi_p$. In each panel we show the mismatch comparisons for a different mode and frame. 
Like the previous example, the mismatch is computed in the frequency band $f \in [20,1024]\ \mathrm{Hz}$ with the LIGO A+ noise curve \cite{LIGO-T1800042}.
The value shown in the Figure is the ratio between the mismatch of the NR waveform and the hybrid and the mismatch of the NR waveform and the TPHM waveform. These values are expected to be similar or less than 1 if no artifacts are introduced during the hybridization procedure. The maximum of the ratios for the modes and frames shown in the plot is 1.008 and the median is 0.937.
Mismatches of the same waveforms are computed for a total mass of $M = 40\ \mathrm{M}_\odot$. Since now the alignment frequency has shifted to lower values, lower values of the mismatch ratio are expected. Indeed, the median drops to 0.753 while the maximum is 1.016.

\subsection{False minima in inertial-frame alignment} \label{ssec:false_minima}

As explained in Sec.~\ref{ssec:hybrids_inertial}, one of the main drawbacks of hybridizing in an inertial frame is the possibility of converging into a value of the shifts that does not correspond to the physically correct alignment. This depends on the metric that is chosen for optimization, but since different waveform models and NR waveforms evolve differently, the notion of a ``true'' alignment is itself ambiguous.

The oscillatory structure inherent to inertial-frame alignment method can obscure the correct alignment, favoring incorrect matches that nevertheless minimize the chosen discrepancy measure for a given system and alignment setup.

% EXAMPLE
To illustrate this issue, we consider an example based on System~B in Table~\ref{tab:discussion_parameters}, where the merger waveform is obtained with \texttt{SEOBNRv5PHM} and the inspiral waveform is obtained with \texttt{IMRPhenomTPHM}.
The alignment is performed within a hybridization window centered at $f_\mathrm{w}$ and spanning $N$ cycles. Optimizing the alignment function defined in Eq.~\eqref{eq:alignment_inertial} reveals typical characteristics of non-linear optimization problems, where different initial guesses converge to different local minima.

\begin{figure*}
    \includegraphics[width=0.24\linewidth]{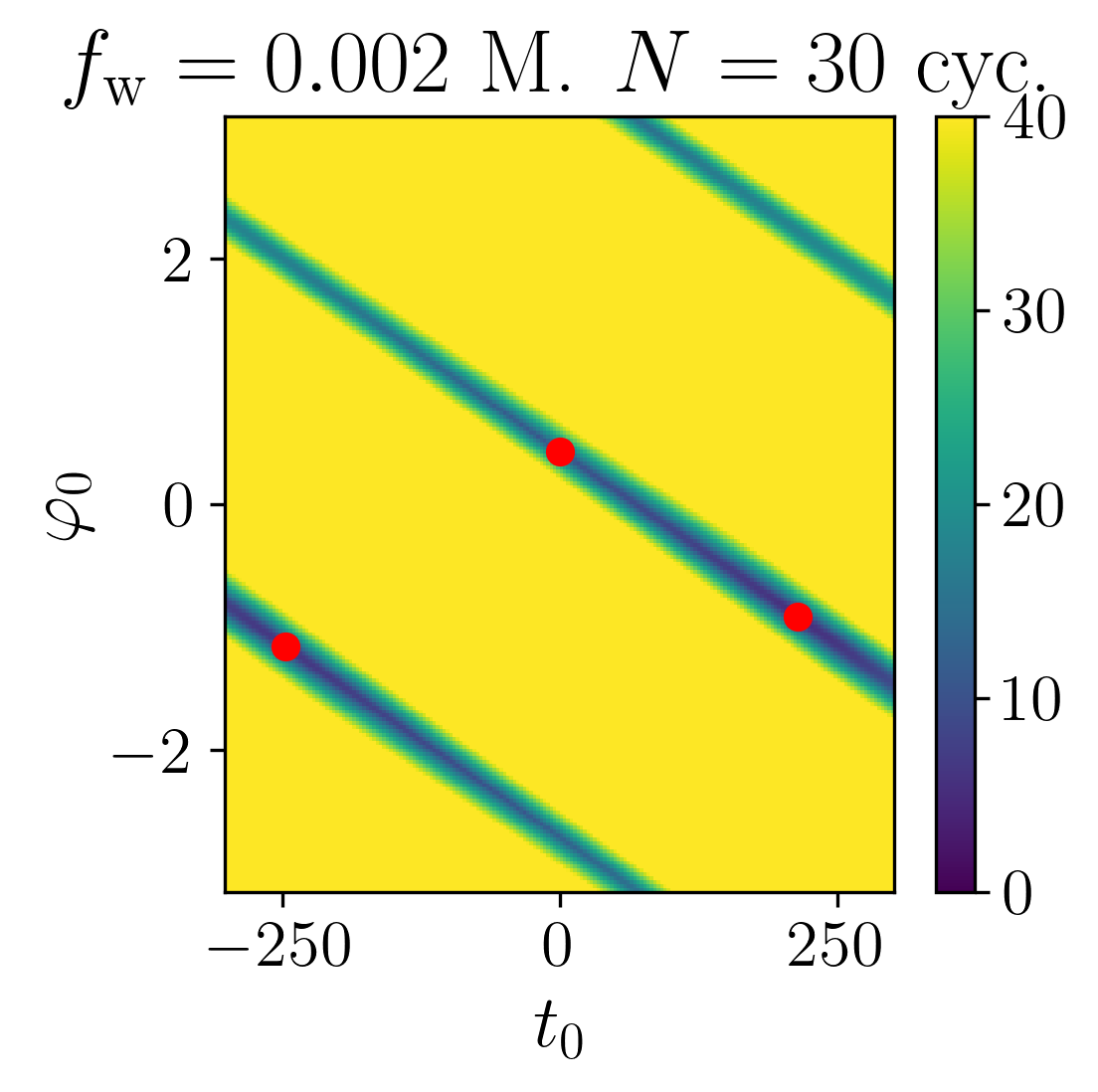}
    \includegraphics[width=0.24\linewidth]{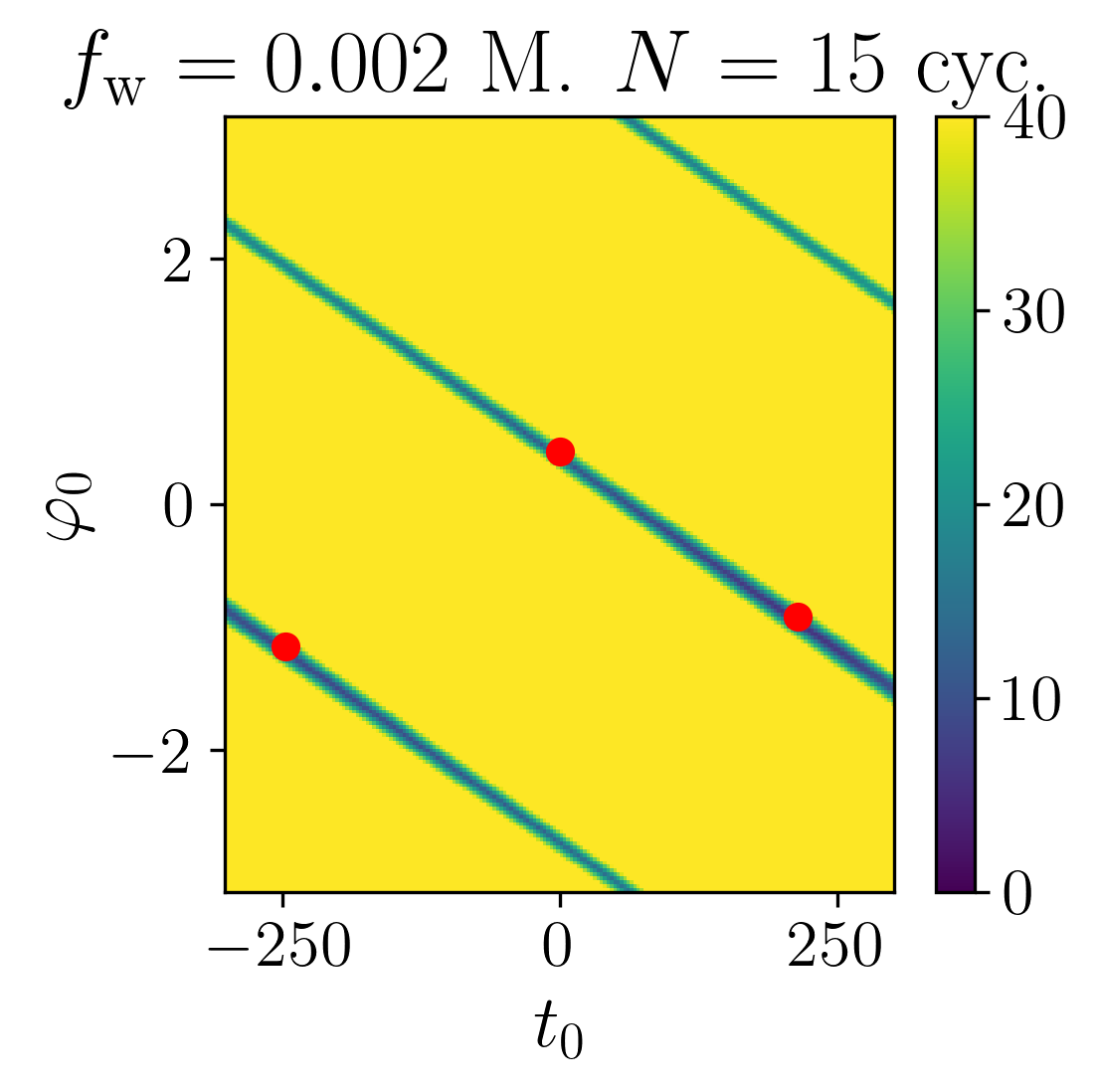}
    \includegraphics[width=0.24\linewidth]{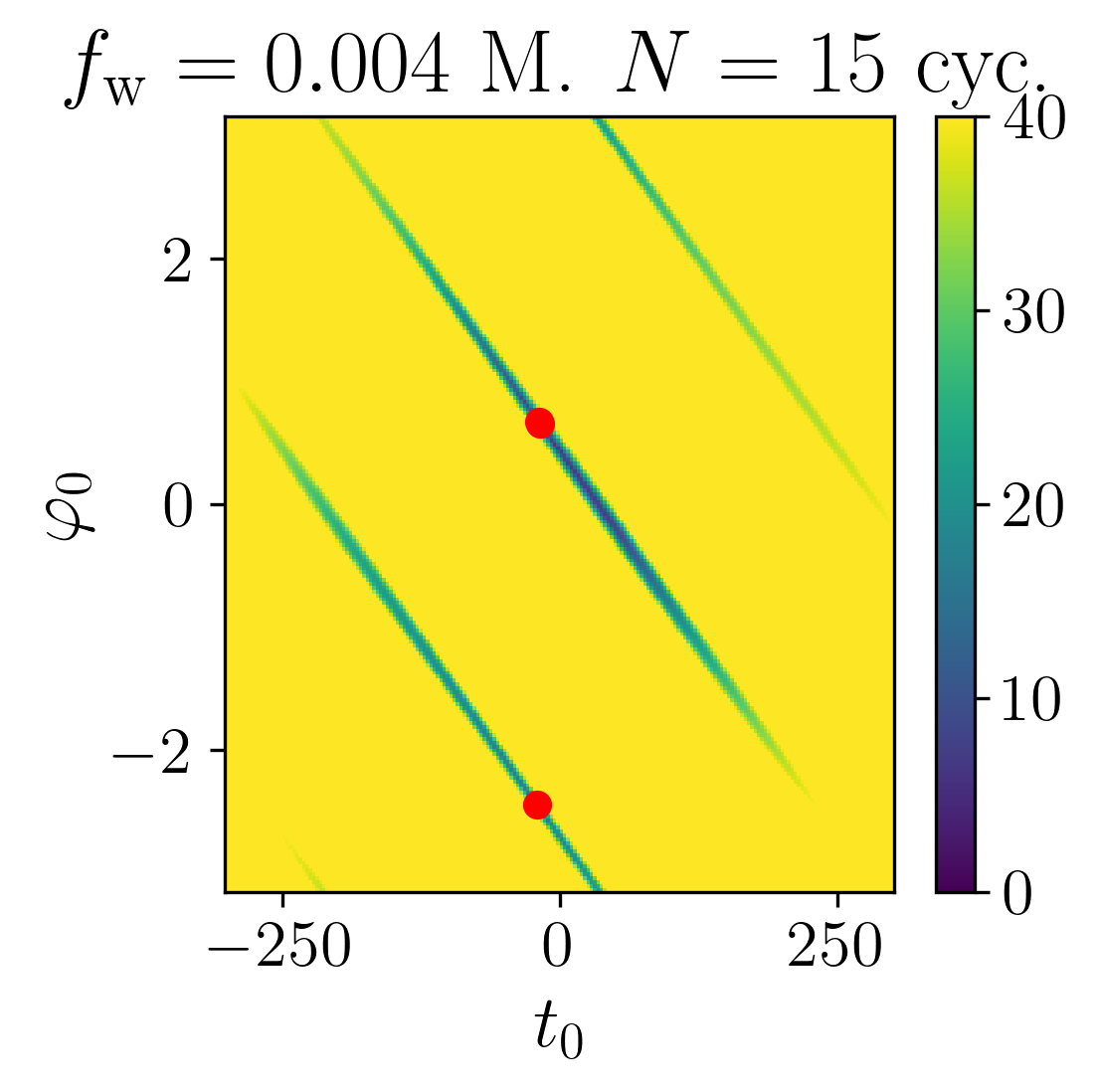}
    \includegraphics[width=0.24\linewidth]{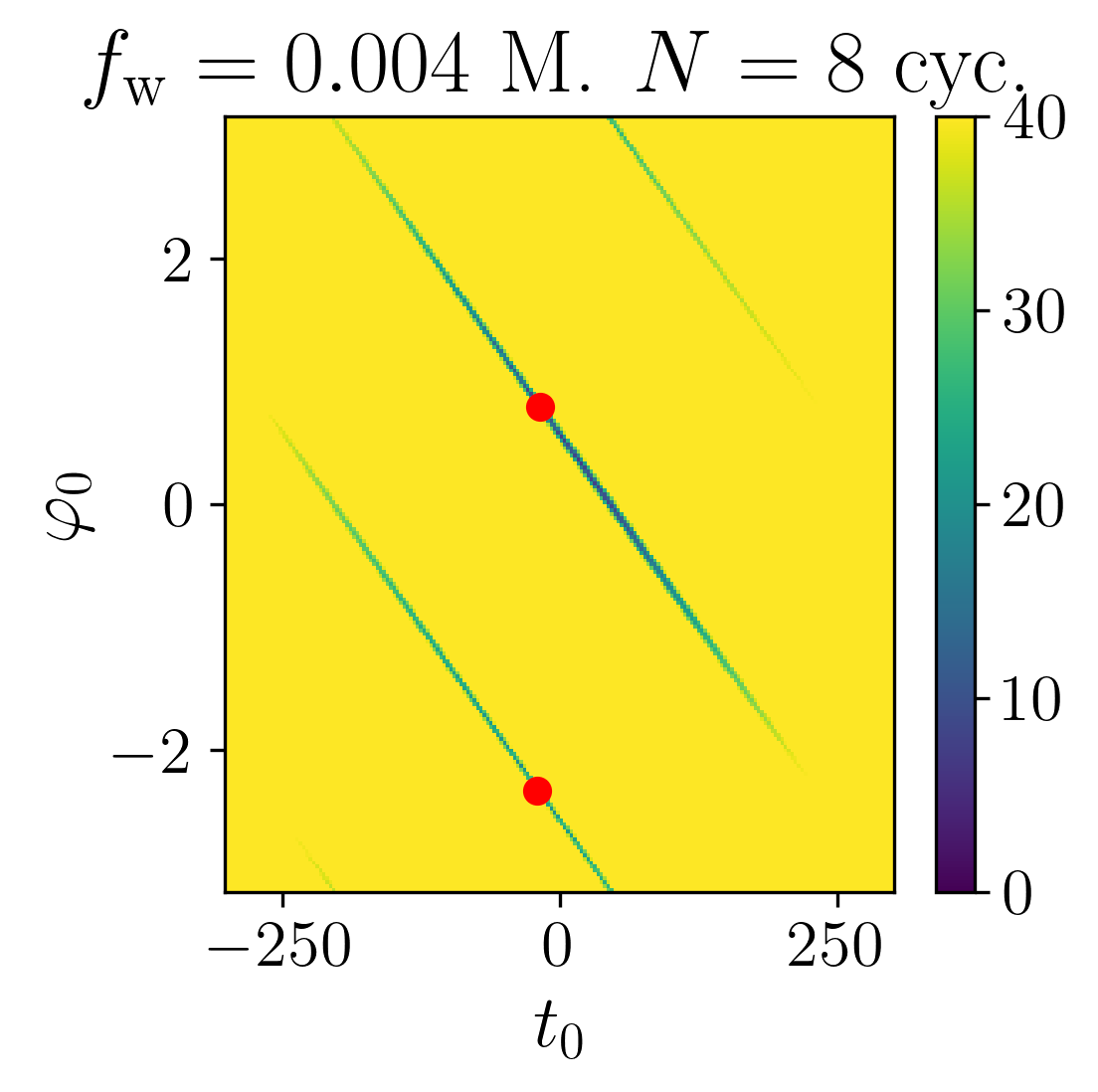}
    \caption{Discrepancy function in Eq.~\eqref{eq:alignment_inertial} as a function of time and phase shifts $(t_0,\,\varphi_0)$, for different alignment window positions and lengths. The central window frequency $f_\mathrm{w}$ and the number of cycles $N$ are indicated. Colors represent the normalized discrepancy, with purple and blue indicating a better alignment. Red dots mark the endpoints of the optimization from various initial guesses. We observe that valleys roughly follow the relation $f_\mathrm{w}t_0-\varphi_0 = \mathrm{const}$, separated by multiples of $\pi$. Later hybridization times and shorter windows reduce degeneracies in local minima.}
    \label{fig:results2}
\end{figure*}

Figure~\ref{fig:results2} displays the discrepancy as a function of $(t_0,\,\varphi_0)$, considering only the (2,2) and (2,1) modes, to avoid complications from incomplete $\ell$-mode content. $\psi_0$ is known to be $\pi/2$ due to the tetrad conventions of the two waveform models involved.

The multiple endpoints (local minima) of the optimization procedure starting from different initial guesses have very close values of the discrepancy function, for very different values of the shifts $(t_0,\varphi_0)$. Therefore, choosing the absolute minimum will give different results for close changes of the alignment window.
This phenomenon reflects the intrinsic nature of the alignment problem rather than a failure of the optimization procedure, and we cannot expect that an improved optimization method will resolve the problem. For some values of the binary parameters, we observe that even a wrong value of $\psi_0$ might lead to optimized values of $t_0,\,\varphi_0$ with lower discrepancy than optimized values with the right $\psi_0$, specially when only a few higher-order multipoles are used in the discrepancy computation.

Additional diagnostics, such as inspecting the phase evolution can help estimate the physically meaningful alignment. However, as seen in Figure~\ref{fig:results2}, at earlier hybridization times, degeneracy is more poorly resolved because the waveform's frequency and amplitude evolve more slowly.

While shorter waveforms are less affected, hybridizing as early as possible remains preferable to maximize the use of information from the merger description. Consequently, this issue is expected to arise frequently in our applications.

The length of the window also affects optimization. As seen in Figure~\ref{fig:results2}, longer windows tend to more local minima and more similar, which hinder the ``true'' value of the shifts. Despite that, short windows are more prone to overfitting.

Ultimately, this challenge makes inertial-frame alignment less suitable for large-scale hybrid production.
Although additional diagnostics could mitigate some of these issues, we instead focus on aligning in the coprecessing frame. The QA frame enables a natural decomposition into amplitude and phase and supports hierarchical alignment strategies, offering a more robust alternative.

\subsection{Parameter estimation studies}

In the development and validation of waveform models, comparing model waveforms with NR simulations is essential. While quantitative measures such as mismatches, amplitude and phase differences provide useful diagnostics, they can be difficult to interpret in terms of their impact on waveform systematics and parameter estimation for real astrophysics signals.

To address this, injection studies are commonly used. In such studies, a waveform generated with the model under investigation is injected mimicking a realistic detector response with specific sky localization, orientation and noise characteristics. A standard PE pipeline is then used to recover the source parameters, typically employing a different waveform model for the recovery. The resulting deviations quantify both statistical and systematic uncertainties.
In the case of precessing systems, additional complications in posterior distributions arise from the time dependence of spin orientations. The choice of reference frequency at which these distributions are evaluated can significantly impact the accuracy of spin parameter recovery.

Here, we use PE to illustrate common issues that appear when hybridizing waveforms from different models.
We have performed several injections and recovered posterior distributions for their parameters with the usual PE pipelines with the waveform model \texttt{IMRPhenomTPHM}. For a choice of NR waveform, we have hybridized with both the \texttt{IMRPhenomTPHM} and the \texttt{SEOBNRv5PHM} models and injected all five waveforms: NR, two models and two hybrids.
Waveforms are generated with all available modes with $\ell \leq 4$ for the NR and model waveforms, and only the common modes when building the hybrids. We use $f_{22,ini} = 10\ \mathrm{Hz}$ as the initial frequency of the model waveforms (and thus the hybrid waveforms) to ensure all modes with $m \leq 4$ are in band as early as $f = 20\ \mathrm{Hz}$. For the shorter NR waveforms the full signal content is not in band until higher frequencies and this may explain why different PE results are obtained.
We have used a total mass of $M = 50\ \mathrm{M}_\odot$ and distance $d = 800\ \mathrm{Mpc}$. We choose these values of total mass and distance to mimic the approximate length and SNR of real events detected by the LVK Collaboration.
Injections were performed for a two detector configuration with L1 and H1 following a LIGO A+ design sensitivity curve \cite{LIGO-T1800042}.
The signal time and sky localisation are identical for all injections and correspond to an apparent altitude of $88.9^\circ$ over the L1 detector site and $63.0^\circ$ over the H1 detector.
The parameter recovery uses \texttt{bilby} \cite{Ashton:2018jfp} with the \texttt{dynesty} sampler \cite{2020MNRAS.493.3132S} with acceptance walk method and common choices of nlive = 1000, naccept = 60.

\begin{figure}
    \includegraphics[width=\columnwidth]{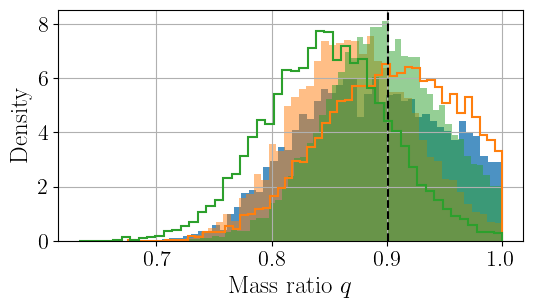}
    \includegraphics[width=\columnwidth]{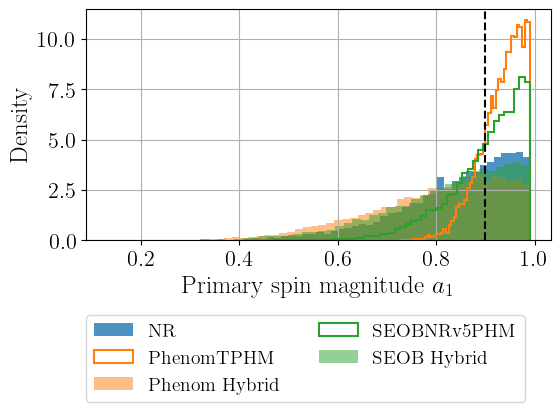}
    \caption{Recovery of the mass ratio (top) and primary spin magnitude (bottom) for injections with the parameters of the NR simulation \texttt{SXS:BBH:0623}, using the \texttt{IMRPhenomTPHM} model for recovery. The true parameters of the injected signal are indicated in a dashed black line.
    Injections are constructed using the NR simulation (blue), an \texttt{IMRPhenomTPHM} waveform (orange, unfilled), an \texttt{SEOBNRv5PHM} waveform (green, unfilled), and the hybrids of the NR waveform with \texttt{IMRPhenomTPHM} waveform (orange, filled) and \texttt{SEOBNRv5PHM} waveform (green, filled).}
    \label{fig:pe1}
\end{figure}

In Fig.~\ref{fig:pe1}, we show the parameter recovery of the five injections corresponding to the NR waveform \texttt{SXS:BBH:0623}. This NR waveform has an initial frequency of $f_{22,ini} = 15\ \mathrm{Hz}$ and is hybridized with an alignment frequency of $f_{22,ref} = 20\ \mathrm{Hz}$.
For all five cases, the mass ratio and primary spin magnitude are recovered within a reasonably low systematic error, even though the posterior distribution is rather broad. Table~\ref{tab:posterior_parameters} provides the recovered parameters.

\begin{figure}
    \includegraphics[width=\columnwidth]{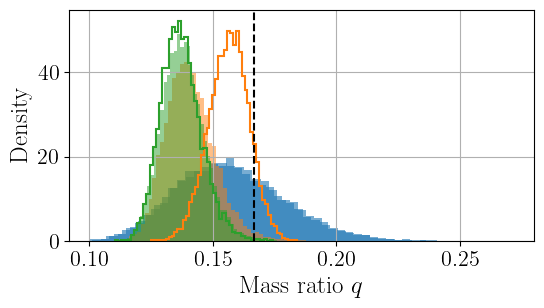}
    \includegraphics[width=\columnwidth]{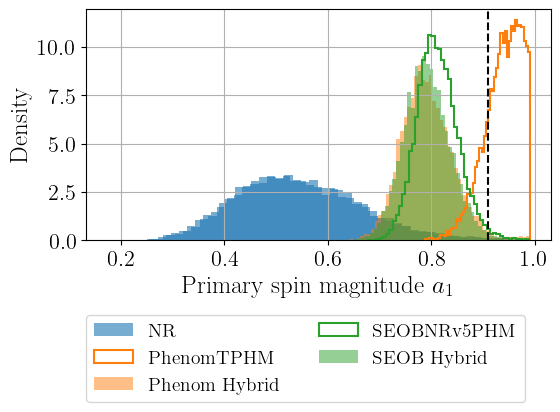}
    \caption{Recovery of the mass ratio (top) and primary spin magnitude (bottom) for injections with the parameters of the NR simulation \texttt{SXS:BBH:0165}, using the \texttt{IMRPhenomTPHM} model for recovery. The legend is identical to Fig.~\ref{fig:pe1}. We observe that the systematic errors are much larger than for the case of \texttt{SXS:BBH:0623}.}
    \label{fig:pe2}
\end{figure}

However, if the NR waveform is too short, especially in the case of low-mass systems where the waveform starts above the detector's sensitivity band, significant biases and increased uncertainties can arise.
These effects are further exacerbated for third-generation or space-based detectors, which are sensitive to lower frequencies.
Short NR waveforms may fail to constrain the true parameters accurately, as their limited duration allows alignment with a wider range of waveforms, regardless of physical accuracy. In such cases, constructing hybrid waveforms by attaching an inspiral segment to the NR portion may help understand these systematics, providing longer signals that capture different posterior distributions.
Figure~\ref{fig:pe2} presents a study using the waveform \texttt{SXS:BBH:0165}, analogous to the analysis in Fig.~\ref{fig:pe1}. For this waveform, the starting frequency of the NR waveform is $f_{22,ini} = 38\ \mathrm{Hz}$ and the alignment interval expands up to $f_{22,ref} = 46\ \mathrm{Hz}$.

In both Figures, the combined network SNR ranges from 30 to 50, with values comparable to real detections.
While the NR waveform \texttt{SXS:BBH:0623} contains $37.5$ orbits, \texttt{SXS:BBH:0165} only contains $6.5$.
This difference is reflected in the results: the injections based on \texttt{SXS:BBH:0623} recover the true parameters better with less bias than the shorter \texttt{SXS:BBH:0165}.
Model-based and hybrid injections also show contrasting behavior: for the longer waveform in Fig.~\ref{fig:pe1}, different model and hybrid injections yield more consistent results, whereas for the shorter signal in Fig.~\ref{fig:pe2}, their disagreement is more pronounced.
A systematic study of biases between models could be carried out for high SNR injections using the Fisher matrix formalism following \cite{Cutler:2007mi}. We leave this for future work.

Together, these figures highlight that direct injections of short NR waveform may struggle to recover accurate parameters. In contrast, hybrid injections --regardless of whether the inspiral model matches the one used for recovery-- often improve the accuracy of the parameter recovery or, at the very least, indicate when the recovery is unreliable.
However, hybridizing closer to merger can introduce significant errors, as gauge differences between the NR and inspiral segments become increasingly important in the late inspiral. This might contribute to hybrid waveforms for these systems also failing to recover the true parameters.

Hybrid injections can also be used to study the influence that different regions of the waveform have on PE. While inspiral, merger and ringdown sections contribute differently to the signal SNR, posterior distributions may pick up features from different regions of the waveform. Studies of hybridized waveforms with different hybridization frequencies may help untangle the contributions of the different sections.

\begin{table*}
\centering
\renewcommand{\arraystretch}{1.1}
\begin{tabular}{
    >{\centering\arraybackslash}m{1.5cm}
    >{\centering\arraybackslash}m{1.5cm}
    >{\centering\arraybackslash}m{2.2cm}
    >{\centering\arraybackslash}m{2.2cm}
    >{\centering\arraybackslash}m{2.2cm}
    >{\centering\arraybackslash}m{2.2cm}
    >{\centering\arraybackslash}m{2.2cm}}
\hline
    \raisebox{.8\height}{\textbf{Parameter}} &
    \raisebox{.8\height}{\textbf{True value}} &
    \raisebox{.8\height}{NR simulation} & 
    \raisebox{.8\height}{\texttt{IMRPhenomTPHM}} & Hybrid \texttt{IMRPhenomTPHM} & \raisebox{.8\height}{\texttt{SEOBNRv5PHM}} & Hybrid \texttt{SEOBNRv5PHM} \\
\hline \hline
    \multicolumn{7}{c}{\texttt{SXS:BBH:0623}} \\ \hline
    $\mathrm{SNR}_\mathrm{H1}$ & & $30.9_{-1.1}^{+1.1}$ & $30.9_{-1.1}^{+1.1}$ & $41.2_{-1.1}^{+1.1}$ & $31.5_{-1.1}^{+1.1}$ & $43.5_{-1.1}^{+1.1}$ \\
    $\mathrm{SNR}_\mathrm{L1}$ & & $34.6_{-1.3}^{+1.2}$ & $34.6_{-1.2}^{+1.2}$ & $46.2_{-1.3}^{+1.2}$ & $35.3_{-1.3}^{+1.3}$ & $48.7_{-1.2}^{+1.2}$ \\
    $\mathcal{M}\ [\mathrm{M}_\odot]$ & $21.73$ & $21.92_{-0.22}^{+0.11}$ & $21.81_{-0.13}^{+0.14}$ & $21.94_{-0.12}^{+0.12}$ & $21.96_{-0.18}^{+0.17}$ & $22.02_{-0.11}^{+0.10}$ \\
    $1/q$ & $0.89$ & $0.89_{-0.10}^{+0.10}$ & $0.90_{-0.10}^{+0.08}$ & $0.87_{-0.08}^{+0.09}$ & $0.85_{-0.09}^{+0.09}$ & $0.90_{-0.08}^{+0.08}$ \\
    $a_1$ & $0.90$ & $0.86_{-0.26}^{+0.11}$ & $0.94_{-0.09}^{+0.04}$ & $0.82_{-0.29}^{+0.15}$ & $0.92_{-0.16}^{+0.07}$ & $0.85_{-0.28}^{+0.13}$ \\
    $a_2$ & $0.90$ & $0.71_{-0.32}^{+0.24}$ & $0.93_{-0.10}^{+0.05}$ & $0.76_{-0.33}^{+0.21}$ & $0.59_{-0.31}^{+0.31}$ & $0.60_{-0.32}^{+0.30}$ \\
    $\chi_\mathrm{eff}$ & & $0.13_{-0.04}^{+0.03}$ & $0.03_{-0.03}^{+0.03}$ & $0.13_{-0.02}^{+0.02}$ & $0.12_{-0.04}^{+0.03}$ & $0.13_{-0.02}^{+0.02}$ \\
    $\chi_\mathrm{p}$ & & $0.83_{-0.17}^{+0.12}$ & $0.93_{-0.08}^{+0.05}$ & $0.82_{-0.18}^{+0.14}$ & $0.82_{-0.18}^{+0.10}$ & $0.82_{-0.20}^{+0.13}$ \\
    $d\ [\mathrm{Mpc}]$ & $800$ & $855_{-36}^{+38}$ & $811_{-32}^{+33}$ & $802_{-77}^{+103}$ & $838_{-71}^{+84}$ & $789_{-75}^{+78}$ \\
\hline
    \multicolumn{7}{c}{\texttt{SXS:BBH:0165}} \\ \hline
    $\mathrm{SNR}_\mathrm{H1}$ & & $20.0_{-1.1}^{+1.1}$ & $24.2_{-1.1}^{+1.1}$ & $24.4_{-1.1}^{+1.1}$ & $25.8_{-1.1}^{+1.1}$ & $24.6_{-1.1}^{+1.1}$ \\
    $\mathrm{SNR}_\mathrm{L1}$ & & $22.4_{-1.3}^{+1.2}$ & $27.1_{-1.2}^{+1.2}$ & $27.4_{-1.2}^{+1.2}$ & $29.0_{-1.2}^{+1.2}$ & $27.6_{-1.2}^{+1.2}$ \\
    $\mathcal{M}\ [\mathrm{M}_\odot]$ & $14.18$ & $13.59_{-0.76}^{+0.81}$ & $14.19_{-0.17}^{+0.15}$ & $14.01_{-0.21}^{+0.21}$ & $14.32_{-0.17}^{+0.16}$ & $14.13_{-0.24}^{+0.24}$ \\
    $1/q$ & $0.17$ & $0.16_{-0.03}^{+0.04}$ & $0.16_{-0.01}^{+0.01}$ & $0.14_{-0.01}^{+0.02}$ & $0.14_{-0.01}^{+0.01}$ & $0.14_{-0.01}^{+0.02}$ \\
    $a_1$ & $0.91$ & $0.53_{-0.17}^{+0.23}$ & $0.94_{-0.07}^{+0.04}$ & $0.79_{-0.07}^{+0.09}$ & $0.81_{-0.06}^{+0.07}$ & $0.79_{-0.07}^{+0.08}$ \\
    $a_2$ & $0.30$ & $0.75_{-0.56}^{+0.22}$ & $0.46_{-0.41}^{+0.46}$ & $0.48_{-0.42}^{+0.46}$ & $0.46_{-0.41}^{+0.47}$ & $0.45_{-0.41}^{+0.47}$ \\
    $\chi_\mathrm{eff}$ & & $-0.47_{-0.19}^{+0.15}$ & $-0.39_{-0.08}^{+0.08}$ & $-0.33_{-0.10}^{+0.09}$ & $-0.23_{-0.08}^{+0.08}$ & $-0.28_{-0.09}^{+0.09}$ \\
    $\chi_\mathrm{p}$ & & $0.14_{-0.04}^{+0.06}$ & $0.83_{-0.05}^{+0.03}$ & $0.68_{-0.05}^{+0.05}$ & $0.76_{-0.05}^{+0.05}$ & $0.72_{-0.05}^{+0.05}$ \\
    $d\ [\mathrm{Mpc}]$ & $800$ & $731_{-167}^{+152}$ & $805_{-43}^{+46}$ & $774_{-48}^{+48}$ & $775_{-39}^{+43}$ & $784_{-47}^{+45}$ \\
\end{tabular}
\caption{Recovered parameters for the two injected systems, using the \texttt{IMRPhenomTPHM} model under the conditions described in the text. The second column lists the true values of the parameters from the NR simulation, while the rest show the recovered values with $90\%$ credible intervals, for each of the five injections. The chirp mass $\mathcal{M}$, inverse mass ratio $1/q$, spin magnitudes $a_1,\,a_2$ and distance $d$ are time-independent, while the effective spin $\chi_\mathrm{eff}$ and effective precessing spin $\chi_p$ are evaluated at the reference frequency specified in the text.}
\label{tab:posterior_parameters}
\end{table*}

\section{Conclusions} \label{sec:conclusions}

% Summary
In this work we have presented a framework to construct accurate hybrid waveforms for generic QC binaries. By aligning the coprecessing frame waveforms and the precessing angle descriptions we obtain hybrid waveforms that are identical to the merger waveforms while extending to lower frequencies.

% Results
We provide a detailed discussion of the challenges intrinsic to precessing waveform alignment together with practical choices that have been made for this work. Precessing systems add three degrees of freedom the aligned-spin systems, which correspond to spatial rotations. For this work, we have neglected the influence that choosing consistent BMS frames would have in correcting discrepancies between gauge-dependent parameters, for a procedure that does ensure consistent BMS frames see \cite{Sun:2024kmv}. A thorough study is left for further work.

The selection of the reference point using the orbit-averaged frequency of the (2,2) mode of the GW in the QA frame relies on Eq.~\eqref{eq:f22-forb}, which might include higher-order corrections. After that, the rotation of spins onto the QA frame relies on the approximation of $\Qhat\approx\Lhat$ and the fixing of $\gamma$ through integrating the frequency evolution from a known point. %Secs.~\ref{ssec:QL}~and~\ref{ssec:orbital_plane} discuss  the approximations made and possible alternatives.
%including the suggestion of an iterative method, is thorough in .
%
Alignment in the inertial frame is discarded due to the lack of robustness of the method. However, we are aware that with different choices of the hybridization interval and the discrepancy function, better behavior can be observed, making alignment feasible. Our choice of hybridizing in the coprecessing frame also leaves room for further study, including but not limited to 1) an adaptive method to choose the hybridization interval, 2) the incorporation of physical priors like orbit counting to estimate shifts like $t_0$ and $\varphi_0$, 3) the use of subdominant modes beyond $(2,1)$ when defining the alignment function, not implemented due to the lack of a model for the (2,0) mode including the relevant memory effects and the complete $\ell=3$ set of modes, and 4) the description of precession using quaternions rather than Euler angles, that would avoid singularities near $\beta=0$.

After describing our strategies, we check the validity of approximating the orbital angular momentum $\Lhat$ with the maximum emission direction $\Qhat$. We also test the degeneracies found when hybridizing in the inertial frame.
Finally, our mismatch and PE recovery studies help illustrate the applications of the hybridization method.

We plan to publish the code used in this work through a \texttt{python} package.

More generally, this process could be extended to account for eccentric waveforms, but critical modifications would be required, due to the non-monotonicity of the GW frequency. In this case, working with orbit-averaged quantities would not be enough to guarantee a correct glueing of the two waveforms.

However, the hybridization strategy presented here is broadly applicable by relying on minimal information about the waveforms and decoupling the alignment of the waveform content from the alignment of reference frames. This approach improves over previous strategies in the broader scope of the method.

% Last paragraph
The systematic production of precessing hybrids will open the door to building catalogs of long NR hybrid waveforms for calibration and validation of waveform models or even parameter estimation studies of real events.
This is specially relevant in the context of third-generation detectors, where long inspirals will be much more common and the need for highly accurate waveform models including precessing effects will become fundamental.

\section*{Acknowledgements}

The authors would like to thank Cecilio García-Quirós for the LSC Publication \& Presentation Committee Review of this manuscript and Antoni Ramos-Buades for comments and suggestions on the paper edition.

We thankfully acknowledge the computer resources at Picasso, the technical support provided by Barcelona Supercomputing Center (BSC) through grant No. AECT-2025-1-0035 from the Red Española Supercomputación (RES), and the Supercomputing and Bioinnovation Center (SCBI) of the University of Malaga for their provision of computational resources and technical support.

This work makes use of the LIGO Algorithms Library \cite{lalsuite} conventions and packages. We use minimization algorithms in \texttt{scipy} \cite{2020NatMe..17..261V}, as well as  \texttt{pyseobnr} \cite{Mihaylov:2023bkc} for the \texttt{SEOBNRv5PHM} model and  \texttt{bilby} \cite{Ashton:2018jfp} for the parameter estimation injection studies.

Joan Llobera Querol is supported through the Conselleria d'Educació i Universitats via an FPI-CAIB doctoral grant FPI\_092\_2022.
Maria de Lluc Planas is supported by the Spanish Ministry of Universities via an FPU doctoral grant (FPU20/05577, EST24/00621).

This work was supported by the Universitat de les Illes Balears (UIB); the Spanish Agencia Estatal de Investigación grants PID2022-138626NB-I00, RED2022-134204-E, RED2022-134411-T, funded by MICIU/AEI/10.13039/501100011033 and the ERDF/EU; and the Comunitat Autònoma de les Illes Balears through the Conselleria d'Educació i Universitats with funds from the European Union - NextGenerationEU/PRTR-C17.I1 (SINCO2022/6719) and from the European Union - European Regional Development Fund (ERDF) (SINCO2022/18146).

\appendix

\section{Fixing the orbital plane} \label{app:orbitalplane}

We discuss here that adding a constant $\varepsilon$ to the Euler angle $\gamma(t)$ is equivalent to shifting the orbital phase $\phiorb$ by $-\varepsilon$.

Geometrically, a constant rotation of the QA frame about the $\zhat$-axis preserves the minimal rotation condition and merely reorients the orbital plane. This redefinition transforms the orbital phase as $\phiorb \to \phiorb-\varepsilon$.
Algebraically, this follows from expressing the QA frame axes (Eqs.~\eqref{eq:xaxis}-\eqref{eq:yaxis}) as
\begin{align}
    \xhat(t) &= \ \, \, \xohat(t)\,\cos{\gamma}+\yohat(t)\,\sin{\gamma} \\
    \yhat(t) &= -       \xohat(t)\,\sin{\gamma}+\yohat(t)\,\cos{\gamma}
\end{align}
where
\begin{equation} \label{eq:xohat-yohat}
    \xohat(t) = \begin{pmatrix}
        \cos{\alpha}\cos{\beta} \\ \sin{\alpha}\sin{\beta} \\ -\sin{\beta}
    \end{pmatrix},\quad
    \yohat(t) = \begin{pmatrix}
        -\sin{\alpha} \\ \cos{\alpha} \\ 0
    \end{pmatrix}
\end{equation}
and the time dependence of $\{\alpha,\beta,\gamma\}$ is implicit.

With this notation, the orbital separation vector in this frame (Eq.~\eqref{eq:n(x,y,phi)}) takes the form
\begin{equation}
    \nhat(t) =  \xohat \cos{(\gamma+\phi)} + \yohat \sin{(\gamma+\phi)},
\end{equation}
which makes it evident that shifting $\gamma(t) \to \gamma(t)+\varepsilon$ induces a compensating shift $\phi(t)\to \phi(t)-\varepsilon$.

This demonstrates that, in non-precessing contexts (either an aligned-spin system or a coprecessing description of a precessing system), the initial orbital phase $\phi_0$ and the constant offset in $\gamma$ are physically degenerate.

\section{Composition of 3D rotations} \label{app:composition}
We remember the definition of the rotation matrix (Eq.~\eqref{eq:xaxis}-\eqref{eq:rotationmatrix}) and we write:
\begin{equation} \tag{\ref{eq:R0}}
    \bm{R}_0 = \bm{R}_\mathrm{B} \cdot {\bm{R}_\mathrm{A}}^\mathrm{T} = \begin{pmatrix}
        r_{xx} & r_{xy} & r_{xz} \\
        r_{yx} & r_{yy} & r_{yz} \\
        r_{zx} & r_{zy} & r_{zz}
    \end{pmatrix}
\end{equation}
with
\begin{widetext}
\begin{equation} \label{eq:explicit_dependence}
\begin{split}
    r_{xx} &= (\cos{\aA} \cos{\aB} \cos{\bA} \cos{\bB} + \sin{\aA} \sin{\aB}) \cos{\dg} - \\
    & \qquad - (\cos{\aA} \cos{\bA} \sin{\aB} - \cos{\aB} \cos{\bB} \sin{\aA}) \sin{\dg} + \cos{\aA} \cos{\aB} \sin{\bA} \sin{\bB} \\ 
    r_{xy} &= (\cos{\aB} \cos{\bA} \cos{\bB} \sin{\aA} - \cos{\aA} \sin{\aB}) \cos{\dg} - \\
    & \qquad - (\cos{\aA} \cos{\aB} \cos{\bB} + \cos{\bA} \sin{\aA} \sin{\aB}) \sin{\dg} + \cos{\aB} \sin{\aA} \sin{\bA} \sin{\bB} \\ 
    r_{xz} &= -\cos{\aB} \cos{\bB} \sin{\bA} \cos{\dg} + \sin{\aB} \sin{\bA} \sin{\dg} + \cos{\aB} \cos{\bA} \sin{\bB} \\
    r_{yx} &= (\cos{\aA} \cos{\bA} \cos{\bB} \sin{\aB} - \cos{\aB} \sin{\aA}) \cos{\dg} + \\
    & \qquad + (\cos{\aA} \cos{\aB} \cos{\bA} + \cos{\bB} \sin{\aA} \sin{\aB}) \sin{\dg} + \cos{\aA} \sin{\aB} \sin{\bA} \sin{\bB} \\
    r_{yy} &= (\cos{\bA} \cos{\bB} \sin{\aA} \sin{\aB} + \cos{\aA} \cos{\aB}) \cos{\dg} - \\
    & \qquad - (-\cos{\aB} \cos{\bA} \sin{\aA} + \cos{\aA} \cos{\bB} \sin{\aB}) \sin{\dg)} + \sin{\aA} \sin{\aB} \sin{\bA} \sin{\bB} \\ 
    r_{yz} &= -\cos{\bB} \sin{\aB} \sin{\bA} \cos{\dg} - \cos{\aB} \sin{\bA} \sin{\dg} + \cos{\bA} \sin{\aB} \sin{\bB} \\
    r_{zx} &= - \cos{\aA} \cos{\bA} \sin{\bB} \cos{\dg} - \sin{\aA} \sin{\bB} \sin{\dg} + \cos{\aA} \cos{\bB} \sin{\bA} \\
    r_{zy} &= - \cos{\bA} \sin{\aA} \sin{\bB} \cos{\dg} + \cos{\aA} \sin{\bB} \sin{\dg} + \cos{\bB} \sin{\aA} \sin{\bA}  \\
    r_{zz} &= \sin{\bA} \sin{\bB} \cos{\dg} + \cos{\bA} \cos{\bB}
\end{split}
\end{equation}
and $\delta\gamma = \gB-\gA$.
\end{widetext}
These expressions make the dependence only on $\delta\gamma$ (and not $\gA$ and $\gB$ individually) explicit.

Equivalently, we can write
\begin{equation} \label{eq:R0_gamma}
\begin{split}
    \bm{R}_0 = &\left[
        \xoBhat \otimes \xoAhat + 
        \yoBhat \otimes \yoAhat \right]\, \cos{\dg}\, + \\
    &+\ \left[
        \yoBhat \otimes \xoAhat -
        \xoBhat \otimes \yoAhat \right]\, \sin{\dg}\, + \\
    &+\ \zBhat \otimes \zAhat,
\end{split}
\end{equation}
where ${\bf{v}} \otimes {\bf{w}} = {\bf{v}} \cdot {\bf{w}}^\mathrm{T}$ and $\xohat$ and $\yohat$ are defined in Eq.~\eqref{eq:xohat-yohat}.

\section{Equivalence of the two methods to obtain the maximum emission direction}
\label{app:ObtainingQ}

The direction of maximum GW emission is estimated in \cite{Schmidt:2010it}, by numerically identifying the axis that maximizes the combined radiation of the $\ell=2$ modes.
This procedure is equivalent to algebraically finding the principal axis of an inertia-like tensor $\mathcal{I}$, defined as:
\begin{equation}
    \mathcal{I} = \iint_{\mathbb{S}^2} \rho(\hat{d_L})\ \mathcal{T}(\hat{d_L})\ \mathrm{d}\Omega
\end{equation}
where $\rho=|h_0|^2$ is the squared amplitude (factoring out the inverse dependence on the luminosity distance), $\mathcal{T}(\hat{d_L})=\hat{d_L}\ \hat{d_L}^\mathrm{T}$, and $\d \Omega = \sin{\theta}\ \d{\theta}\ \d \varphi$.
The function $\rho(\theta,\varphi)$ can be expressed explicitly using the spin-weighted spherical harmonic decomposition of the waveform, allowing $\mathcal{I}(\theta,\varphi)$ to be computed in closed form.

An equivalent algebraic method to compute the principal axes of radiation has been introduced in \cite{OShaughnessy:2011pmr}, inspired by quantum mechanics.
It constructs a tensor $\langle \mathcal{L}_{(ab)} \rangle$ as the average of products of the rotation group generator operators $L_k$, applied to $\psi_4$. Diagonalizing this tensor yields its principal axes, with the dominant emission direction given by the eigenvector corresponding to the largest eigenvalue.
After algebraic manipulation, one finds that $\mathcal{I}$ and $\langle \mathcal{L}_{(ab)} \rangle$ differ only  by a normalization and a possible multiple of the identity tensor, which do not affect the determination of the principal axes.

%\vspace{0.1in}
%\vfil
\let\c\Originalcdefinition %
\let\d\Originalddefinition %
\let\i\Originalidefinition
\bibliography{bibliography}

\end{document}